\documentclass[%
 reprint,
nofootinbib,
amsmath,
amssymb,
aps,
prc,
floatfix,
]{revtex4-2}

\usepackage{latexsym}
\newcommand{\bqa}{\begin{eqnarray}}
\newcommand{\eqa}{\end{eqnarray}}
\usepackage{amsmath}
\usepackage{graphicx}
\usepackage{dcolumn}
\usepackage{bm}
\usepackage[utf8x]{inputenc}
\usepackage{hyperref}
\hypersetup{linkcolor=blue}

\begin{document}

\preprint{APS/123-QED}

\title{Using the non-hydrodynamic mode to study the onset of hydrodynamic behavior in ultraperipheral symmetric nuclear collisions}
\author{Nikhil Hatwar}
\email{nikhil.hatwar@gmail.com}
\affiliation{Department of Physics, Birla Institute of Technology and Science, Pilani, Rajasthan, India}
\author{M. Mishra}
\email{madhukar@pilani.bits-pilani.ac.in}
\affiliation{Department of Physics, Birla Institute of Technology and Science, Pilani, Rajasthan, India}

\date{\today}

\begin{abstract}
With the attempts of extending the hydrodynamic framework of heavy-ion collision to proton-proton and other small and low energy systems, we are confronted with the question of how small the system can get and still be safely modelled as a fluid. One of the transport coefficients required in the $2^{nd}$ order relativistic viscous hydrodynamics is the shear relaxation time, inclusion of which solves the causality violation problem in the Navier-Stokes equation. In phenomenological studies this coefficient has been taken as a constant and much attention has gone into finding and fixing the shear viscosity to entropy density ratio, $\eta/s$. This transport coefficient also happens to control the non-hydrodynamic mode of the out-of-equilibrium hydrodynamics theory. It has been predicted that for decreasing system size, observables become sensitive to variation in shear relaxation time as a result of increasing dominance of non-hydrodynamic mode, which could potentially indicate breakdown of hydrodynamics. In this study, we try to test this prediction in the peripheral Pb-Pb collisions at $2.76$ TeV and Au-Au collisions at $200$ GeV, with IPGlasma initial condition and $(2+1)-$Dimensional viscous hydrodynamics. We find that elliptic flow does show adequate sensitivity to variation in relaxation time for decreasing system size. The multiplicity rapidity density limit for applicability of hydrodynamics is found to be around $dN/dy\approx10$, with the possibility of refinement in this value given a way to improve the centrality resolution in experimental data for referencing in peripheral collisions.
\end{abstract}

\maketitle
\section{\label{sec:intro} Introduction}

The fact that baryons have internal structure directly leads to the notion that a bulk medium of sub-nucleonic degrees of freedom should exist~\cite{QGP_idea1,QGP_idea2}. An energy density of about $0.7$ GeV/fm$^3$ is required to free up quarks from the nucleons~\cite{big_picture,ed_for_QGP}. We now have convincing signs from experiments at the BNL Relativistic Heavy Ion Collider (RHIC) and the CERN Large Hadron Collider(LHC) that indicate a deconfined state of quarks and gluons called quark-gluon plasma (QGP) is formed for a sufficient distinguishable duration. Low-order hydrodynamic constitutive relations apparently explain the experimental observables of such a dynamic system quite well, even though there is a sizable pressure anisotropy. This applicability of low-order hydrodynamics has been referred to as \textit{hydrodynamization},\footnote{In this study, we will refer to the applicability of low-order hydrodynamics as "hydrodynamization", in accordance with its definition in Ref. ~\cite{hydrodynimization1}.}, to distinguish it from local thermalization ~\cite{hydrodynimization1,hydrodynimization2}. Experimental confirmation of strangeness enhancement~\cite{2000observing}, elliptic flow~\cite{2001elliptic} and jet quenching~\cite{jet2001,jet2003} as the early indicators was subsequently followed by confirmation of other signatures like quarkonia suppression. Efforts now are directed towards quantitatively fixing the boundaries of various regions of Quantum Chromodynamics(QCD) phase diagram~\cite{BES_2013} and deducing the properties of QGP~\cite{Nature_Bass}.

There are challenges involved in analytically solving non-perturbative QCD making the proof of deconfinement intractable~\cite{wu1991relativistic}. Hence the progress in modelling a medium of quarks and gluons from first principles has been limited. Lattice QCD, even though computationally intensive, has been of help in understanding deconfinement, and other low density phenomena where the numerical sign problem does not affect the calculations~\cite{2018latticeReview,bazavov2014equation}. For now, phenomenological models aided by lattice QCD seem to be the right approach in modelling such a complex system. The use of hydrodynamics in modelling the transient QGP stage has been quite surprising~\cite{2008hydro}. However, hydrodynamics as an effective theory for heavy-ion collisions, has evolved tremendously, especially in the last two decades.
For an in-depth review of the hydrodynamics in heavy-ion collisions, please lookup Refs. ~\cite{2010roma,kovtun2012lectures,2016Ulrich,jaiswal2016,florkowski2018new,blaizot2020emergence,2017roma_app_hydro}.
Apart from the traditional conversation equation approach, hydrodynamics can also be derived as a microscopic theory in the limit e.g., starting from kinetic theory or any QFT like QCD provided its dynamics show a quasi-universality at a large time scale~\cite{florkowski2018new}. This microscopic theory approach also helps in fixing the transport coefficients of the theory~\cite{kovtun2012lectures}.

The energy momentum tensor for such a theory in a non-equilibrium state is decomposed as;
\begin{equation}
T^{\mu\nu} =  \langle \hat{T^{\mu\nu}} \rangle_{eq} + \delta \langle \hat{ T^{\mu\nu}} \rangle,
\end{equation}
where the first and second term represents the equilibrium state of $T^{\mu\nu}$ and the deviation from the equilibrium, respectively. Under linear response theory, the second term can be expanded as;
\begin{equation}
 \delta \langle \hat{T^{\mu\nu}} \rangle (x) = -\frac{1}{2} \int d^4y G^{\mu\nu,\alpha\beta}_R (x^0-y^0,\textbf{x}-\textbf{y})  \delta g_{\alpha\beta}(y),
\end{equation}
where, $G^{\mu\nu,\alpha\beta}_R (x^0-y^0,\textbf{x}-\textbf{y})$ is the retarded $2-$point correlator of $T^{\mu\nu}$.
And $\delta g_{\alpha\beta}(y)$ is a small perturbing term added to flat space-time metric.
This correlator when expressed in the Fourier space [$G^{\mu\nu,\alpha\beta}_R (\omega,\textbf{k})$], where $\omega$ is the angular frequency and, $\textbf{k}$ is the momentum, has singularities. The solution of the $\delta \langle \hat{T^{\mu\nu}} \rangle (x)$ integral at late times has a contribution in terms of complex singular frequency in the $\omega$-plane;
\begin{equation}
\omega_{sing} = \omega_h + i \; \omega_{nh}
\end{equation}
where, $\omega_h$ is the real part of frequency at singularity corresponding to excitation of equilibrium plasma, also called \textit{hydrodynamic mode frequency}. $\omega_{nh}$ is termed as transient mode or \textit{non-hydrodynamic mode frequency} and is associated with the dissipative effects. The transient mode is responsible for disruption of  hydrodynamization process and is controlled with the \textit{relaxation time} parameter which sets the duration for which viscous effects remain active. These are called the \textit{quasi-normal modes} of out-of-equilibrium hydrodynamics, analogous to the \textit{normal modes} of oscillatory systems in classical mechanics.

Right after the collision of heavy-ions, we have a non-equilibrium system of partons for upto $1$ fm/c.
The fact that applying low-order hydrodynamics does not require local thermalization or even pressure isotropy to show agreement with the measurements~\cite{roma2017_hydro}, had been puzzling, until we discovered that this evolution leads to an attractor ~\cite{2013hydro_Heller1,2015hydro_Heller2,romatschke2017relativistic,hydro_aleksi2020}. This attractor guides the system evolution to a late time universal trajectory even if initiated with a varied set of starting conditions~\cite{universality}.

The framework of hydrodynamics with initial conditions, followed by a hadron after-burner, has been quite successfully used to explain experimental data obtained from a wide range of systems~\cite{2020gamut,2016Roma_pp}. From most central to ultra peripheral collisions, the system size decreases monotonically. For a constant collisional energy, there should to be a system size below which the QGP droplet will cease to \textit{hydrodynamize}~\cite{nagle2018small}. Aleksi Kurkela \textit{et al.},~\cite{kurkela2019flow,kurkela2019opacity} has performed a flow analysis with kinetic theory leading to hydrodynamization, through a dimensionless physical quantity called opacity($\hat{\gamma}$) -- a measure of transverse system size in units of the mean free path. As the opacity varies from $0$ to $5$, the system goes through $3$ stages in this order:(a) non-QGP (particle-like) stage, (b) intermediate transition stage and (c) QGP (hydro-like) stage. Ulrich Heinz and Moreland~\cite{heinz2019_how_the_heck} have emphasized considering the multiplicity rapidity density of charged particle -- $dN/dy$ along with HBT radii to quantify the smallest QGP size. According to Romatschke~\cite{2017roma_app_hydro}, the large $p_T$ regime of flow is due to non-hydrodynamic mode and this mode can be studied through the relaxation time approach. He suggested that large deviation of elliptic flow ($v_2$) for a variation in shear relaxation time for lowering multiplicity could potentially indicate breakdown of low-order hydrodynamics. The last two of the above studies came to the conclusion that this limit should be around or below $dN_{ch}/dy\approx 2$.

The role of relaxation time has been previously analyzed for different settings in hydrodynamics studies ~\cite{old_relax_study1,old_relax_study2,old_relax_study3,old_relax_study4,old_relax_study5,old_relax_study6}, including spatial and momentum eccentricity, entropy and elliptic flow for varying relaxation times. However the primary focus of these studies was to find the range of $\tau_\pi$ and other second order transport coefficients for which the observables were insensitive, which in turn meant that the magnitude of the second order gradient terms are smaller in comparison to those of first order gradient. In the present work, we check the sensitivity of observables to shear relaxation time in ultra peripheral collision systems to test the breakdown of low-order hydrodynamics. In Sec. \ref{method} we discuss the framework of the model used. In Sec. \ref{parameters}, we state the initial condition and input parameters involved in the model. Sec. \ref{fix_parameters} describes the observables obtained along with the experimental results in order to fix the centrality related parameters. In Sec. \ref{sec:results}, we present results of elliptic flow as a function of transverse momentum and multiplicity rapidity density. And in Sec. \ref{sec:conclu}, inferences are drawn based on results obtained along with the possible improvement to this work.

\begin{figure}
\begin{center}
\includegraphics[scale=0.45]{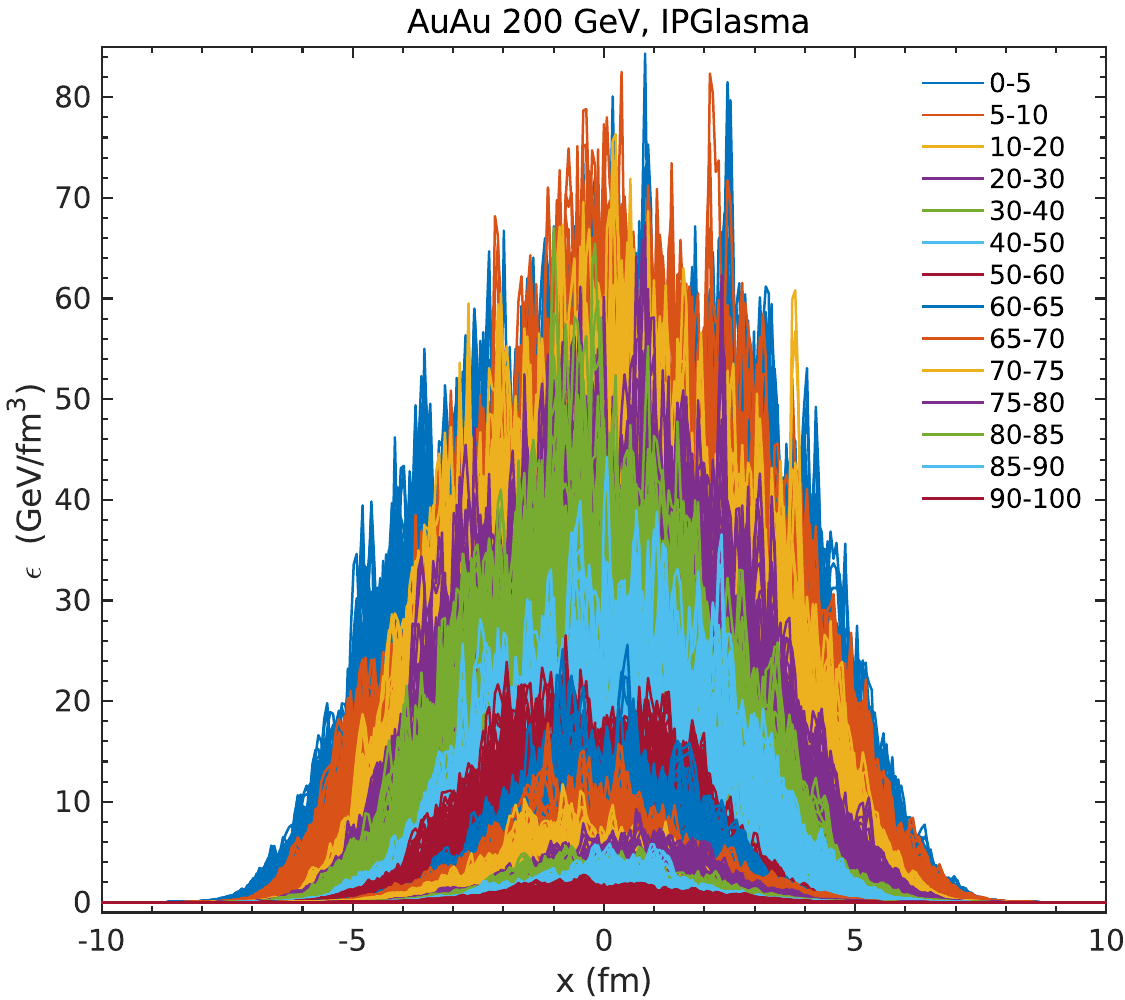}
\end{center}
\caption{Energy density distribution as a function of transverse coordinate at $\tau=0.6$ fm, midrapidity for $14$ centrality classes of Au-Au IPGlasma runs at $200$ GeV. The distribution for each centrality class has been superimposed for $400$ IPGlasma events with different nucleon positions to account for event-by-event fluctuations.}
\label{fig:ed}
\end{figure}

\begin{figure*}
\begin{center}
\includegraphics[scale=0.37]{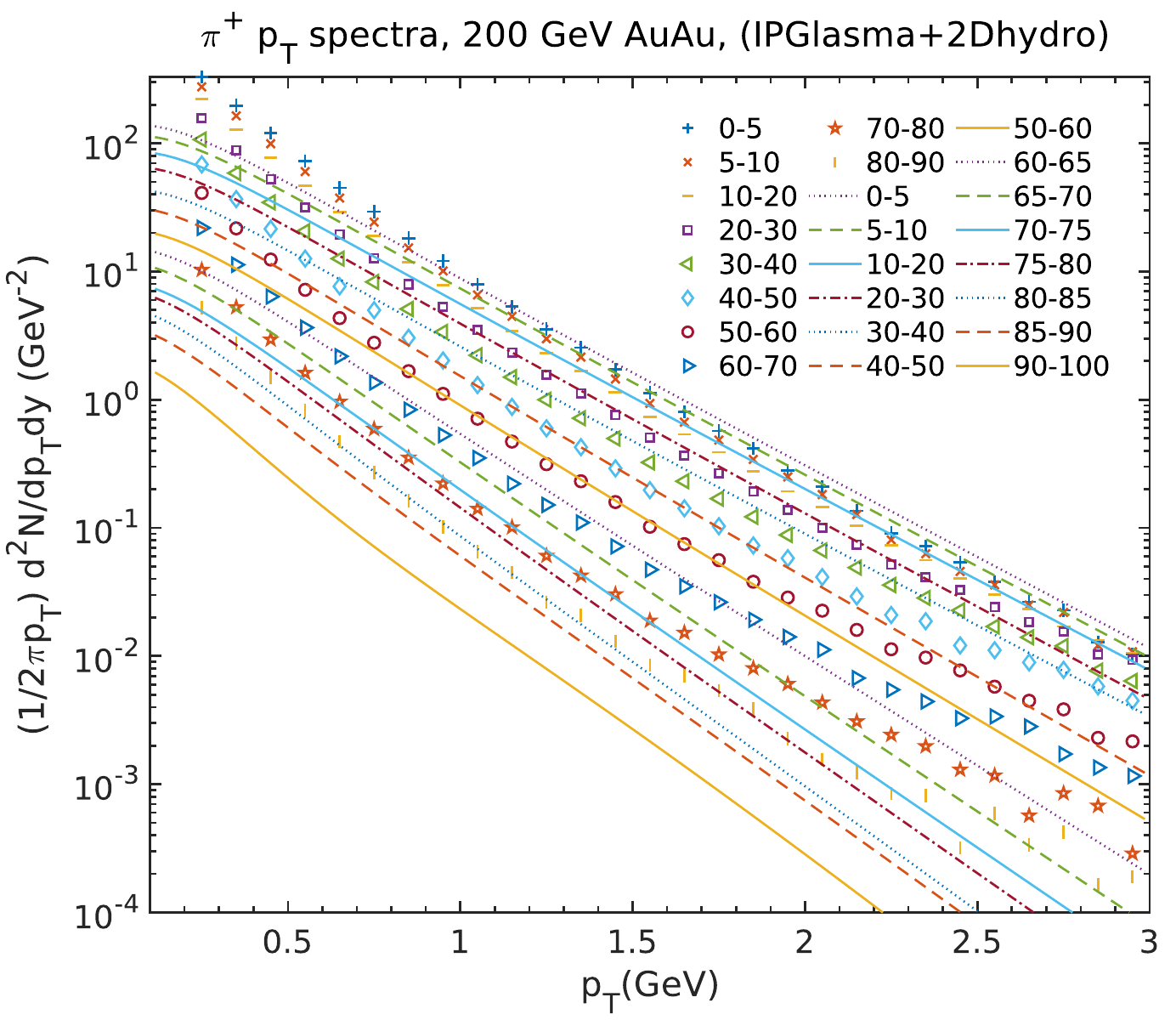}
\includegraphics[scale=0.37]{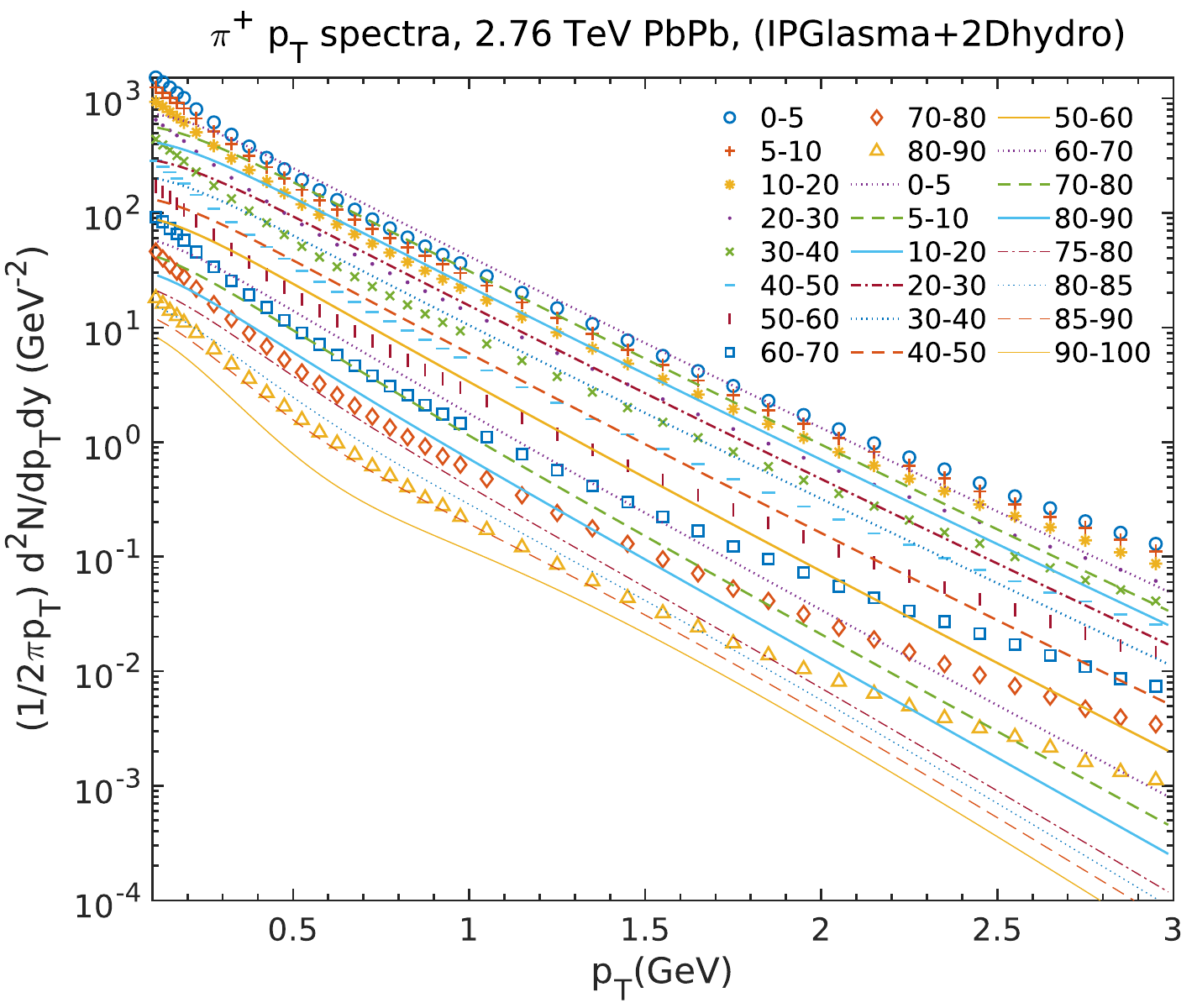}
\end{center}
\caption{Pion($\pi^{+}$) $p_T$-spectra generated (lines) for Au-Au at $200$ GeV (left) and Pb-Pb at $2.76$ TeV (right) for mentioned centrality classes compared with the corresponding PHENIX \cite{AuAu_pT_spec} and ALICE experimental results ~\cite{expt_adam2016_pT_spec} (symbols).}
\label{fig:pT_spectra}
\end{figure*}

\begin{figure}
\includegraphics[scale=0.32]{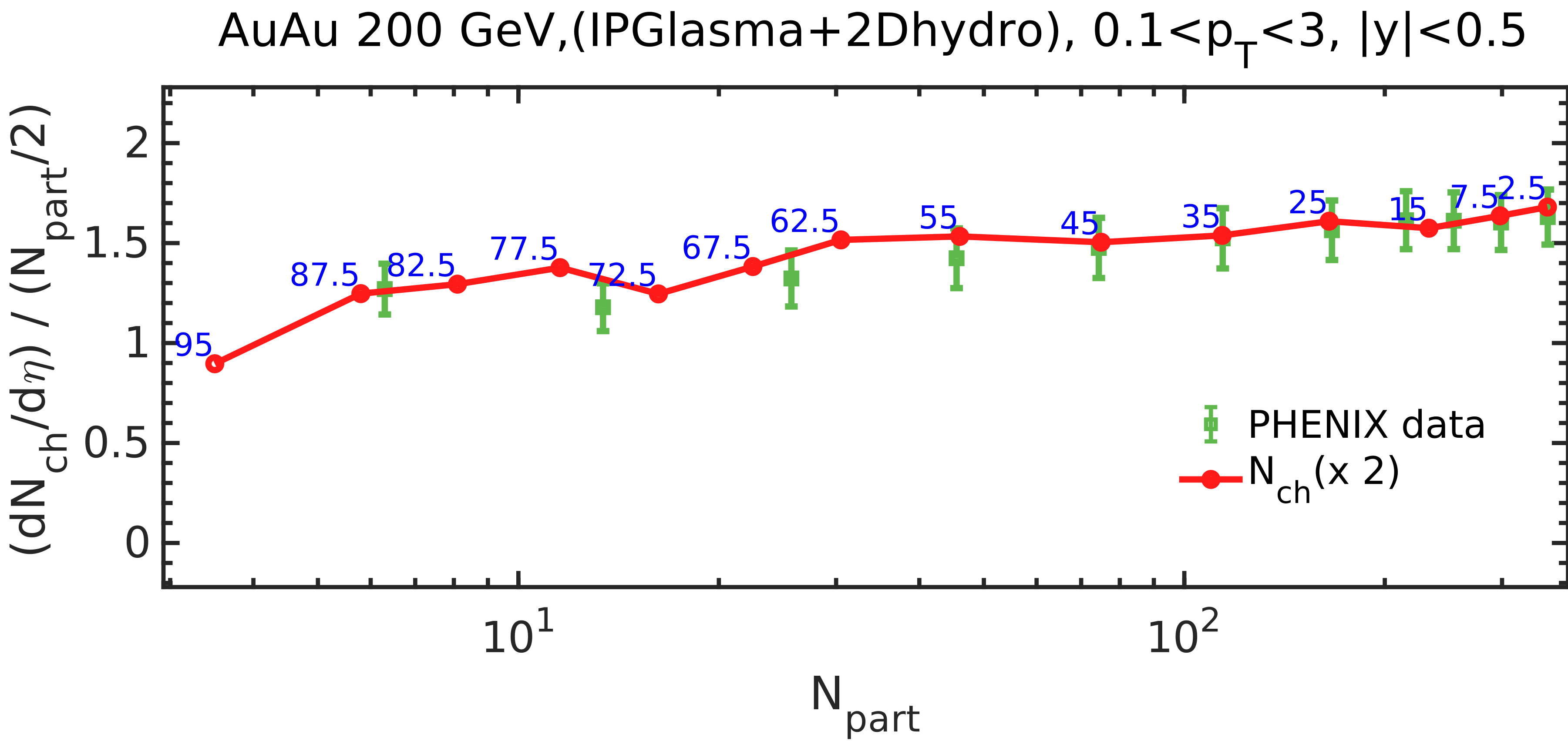}
\includegraphics[scale=0.30]{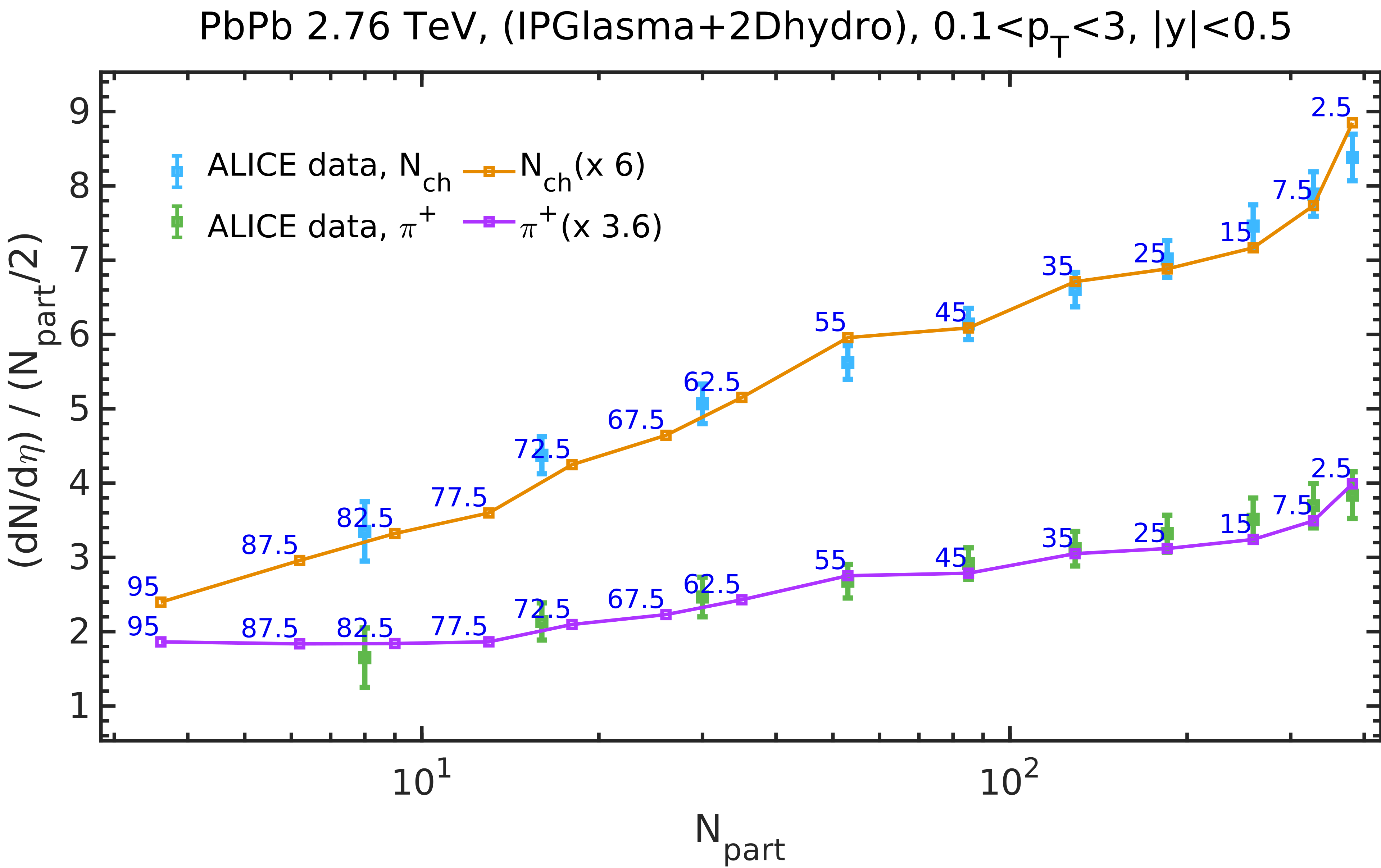}
\caption{Charged particle multiplicity rapidity spectra generated (lines) for Au-Au $200$ GeV (above) and Pb-Pb $2.76$ TeV (below) as a function of number of participants compared with corresponding PHENIX\cite{AuAu_pT_spec} and ALICE\cite{PbPb_expt_dNdy_Npart,PbPbExpt_Nch_dNdy} experimental data (errorbars). The generated data points are labelled with the midpoint of the centrality range in blue color.}
\label{fig:dNdy_Npart}
\end{figure}

\section{\label{method} Formalism}

Hydrodynamics is the collective dynamical evolution of a suitably sized bulk medium adhering to the system's symmetries. For the relativistic case, the conservation laws take the form, $\partial_{\mu} T^{\mu\nu}=0$ for energy–momentum tensor and $\partial_{\mu} N^\mu=0$ for conserved charge. The local values of temperature, $T(x)$, fluid velocity, $u_\mu(x)$ and chemical potential, $\mu(x)$ are chosen as hydrodynamic variables. For ultra-relativistic collisions, where a negligible amount of participating nucleons survive, the conservation equation for baryon number ($\partial_{\mu} N^\mu=0$) can be ignored. The energy-momentum tensor can be decomposed as~\cite{1940Eckart};
\begin{equation}
T^{\mu\nu}=\epsilon u^\mu u^\nu + \Delta^{\mu\nu} P+ ( w^\mu u^\nu + w^\nu u^\mu) + \Pi^{\mu\nu}.
\end{equation}
Here, $\epsilon$(energy density) and $P$(pressure) are scalar coefficients. $w^\mu$ represents the transverse vector coefficient.
$\Delta^{\mu\nu} \equiv g^{\mu\nu} +  u^\mu u^\nu$ is the projector operator orthogonal to the fluid velocity($u^\mu$) and $g^{\mu\nu}$ is the space-time metric. The above expression without the $\Pi^{\mu\nu}$ term corresponds to $0^{th}$ order ideal hydrodynamics. The $\Pi^{\mu\nu}$ tensor is introduced to account for the dissipative effects and is further decomposed as:
\begin{equation}
\label{dissi_tensor}
\Pi^{\mu\nu}= \pi^{\mu\nu} +  \Delta^{\mu\nu} \Pi.
\end{equation}
$\Pi$ and $\pi^{\mu\nu}$ are the bulk and shear part of the viscous stress tensor. The form of the shear stress tensor($\pi^{\mu\nu}$) and bulk pressure($\Pi$) are set up in accordance with the covariant form of the second law of thermodynamics~\cite{2010roma}. When we set entropy $4-$current expression  as $s^\mu = s u^\mu$, where $s$ is entropy density, we get;
\begin{equation}
\label{first_order}
\pi^{\mu\nu} = \eta \sigma^{\mu\nu} \qquad \textnormal{and}  \qquad \Pi = \zeta \; \partial_\mu u^\mu,
\end{equation}
where $\eta$(shear viscosity) and $\zeta$(bulk viscosity) are the transport coefficients. $\sigma^{\mu\nu}$(shear tensor) is a traceless, transverse and symmetric tensor.
This form of $\pi^{\mu\nu}$ and $\Pi$ leads to the $1^{st}$ order, Navier–Stokes theory.
When we introduce perturbations in energy density and fluid velocity, and evolve them, the diffusion speed obtained from the dispersion relation has a form that can increase arbitrarily.
This theoretical formulation cannot be considered as a satisfactory one, if it violates causality. It turns out that if the term ($-\tau_\pi u^\alpha \partial_\alpha \pi^{\mu\nu}$) is added in the expression of $\pi^{\mu\nu}$ above, the resulting diffusion speed stays below the speed of light. The coefficient of this newly added term, $\tau_\pi$ is called \textit{relaxation time}.
But this is still a makeshift way to restore causality in the system. A good $2^{nd}$ order viscous hydrodynamics theory at the very least should reduce to the Navier–Stokes equation in the limit of long wavelengths, and must show causal signal propagation.

Müller~\cite{muller1967}, Israel and Stewart~\cite{israel1976,israel1979}(MIS) suggested modification of the entropy $4-$current expression used above to include the following term with a viscous stress tensor:

\begin{equation}
s^\mu = s u^\mu - \frac{\beta_0}{2T} u^\mu \Pi^2 - \frac{\beta_2}{2T} u^\mu \pi_{\alpha\beta}\pi^{\alpha\beta} + \mathcal{O}(\Pi^3)
\end{equation}
where $\beta_0$ and $\beta_2$ are scalar coefficients.
When we use this entropy $4-$current in covariant $2^{nd}$ law of thermodynamics, the dissipative terms of energy momentum tensor take the following forms~\cite{2010roma}:
\newcommand{\hydro}{${}_{\alpha}u_{\beta}$}

\bqa
\pi_{\alpha\beta} =&& \; \eta \Bigg( \nabla_{\langle\alpha} u_{\beta\rangle}  - \pi_{\alpha\beta} T u^\mu \partial_\mu \Bigl(\frac{\beta_2}{T} \Bigr) - 2\beta_2  u^\mu  \partial_\mu   \pi_{\alpha\beta}  \nonumber\\ &&  - \;\; \beta_2 \pi_{\alpha\beta}  \partial_\mu u^\mu   \Bigg)
\eqa
\bqa
\Pi = && \; \zeta \Bigg(  \nabla_\alpha u^\alpha - \frac{1}{2}\Pi \; T \; u^\mu \partial_\mu \Big( \frac{\beta_0}{T} \Big) -  \beta_0   u^\mu \partial_\mu \Pi \nonumber\\ && - \; \; \frac{1}{2} \beta_0  \Pi \partial_\mu u^\mu  \Bigg)
\eqa

Where, $\nabla^\mu = \Delta^{\alpha\mu} \partial_\alpha$ and $\nabla_{\langle {}_{\alpha}u_{\beta} \rangle}$ is a symbol to represent traceless symmetrization of $\nabla_\alpha u^\beta$.
A perturbative analysis with these newly obtained expressions leads to an inherently causal system. There are a few variants of this theory~\cite{DNMR2010}, depending on how many terms are kept in $\pi^{\mu\nu}$ and $\Pi$ expression.
The viscous hydrodynamics code used for this study is based on MIS theory. BRSSS theory~\cite{BRSSS2008} is a more comprehensive version of MIS hydrodynamics. A few $3^{rd}$ order versions have also been worked up~\cite{2013Jaiswal,diles2020third}.

\begin{figure*}
\begin{center}
\includegraphics[scale=0.29]{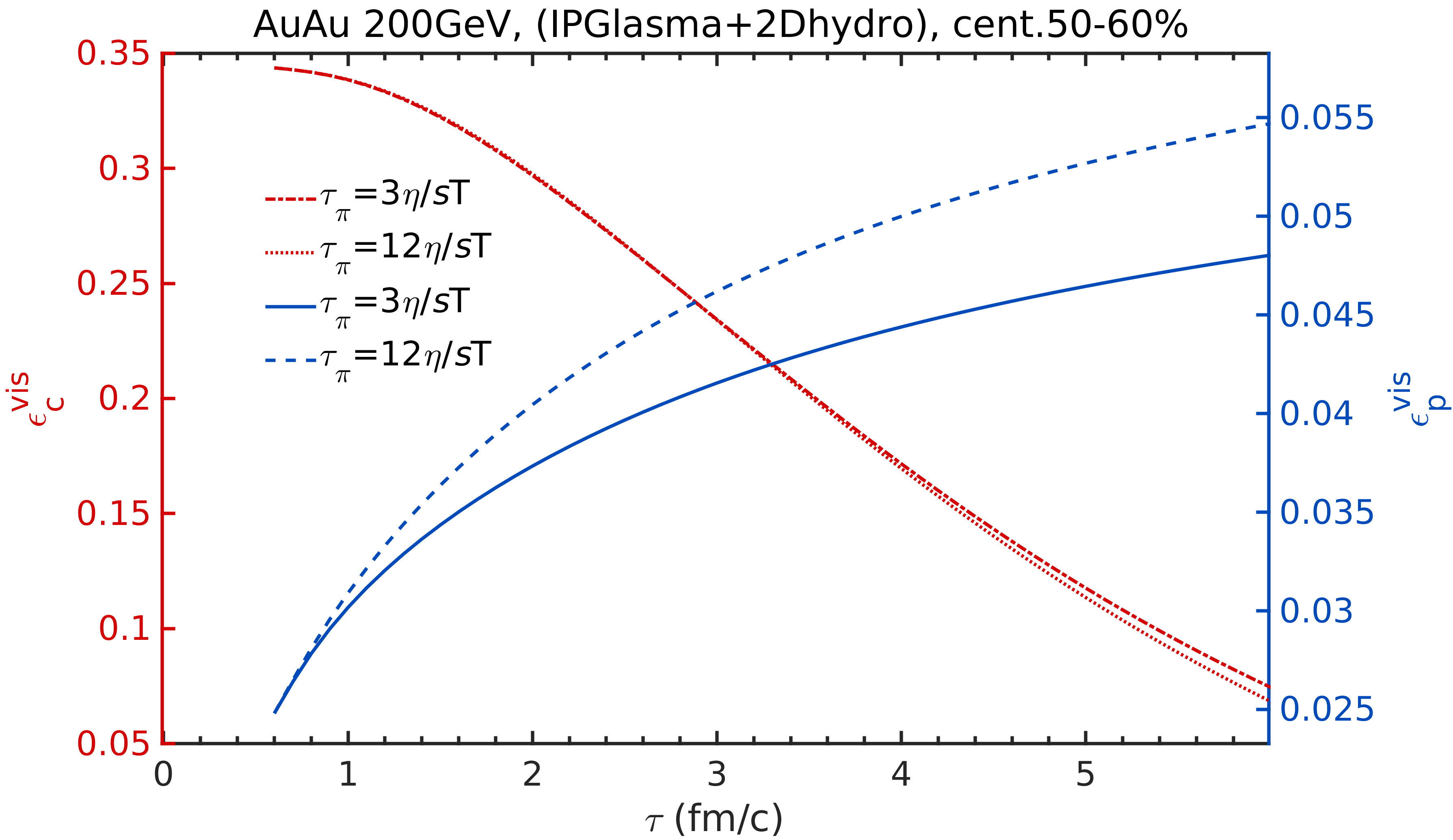}
\includegraphics[scale=0.29]{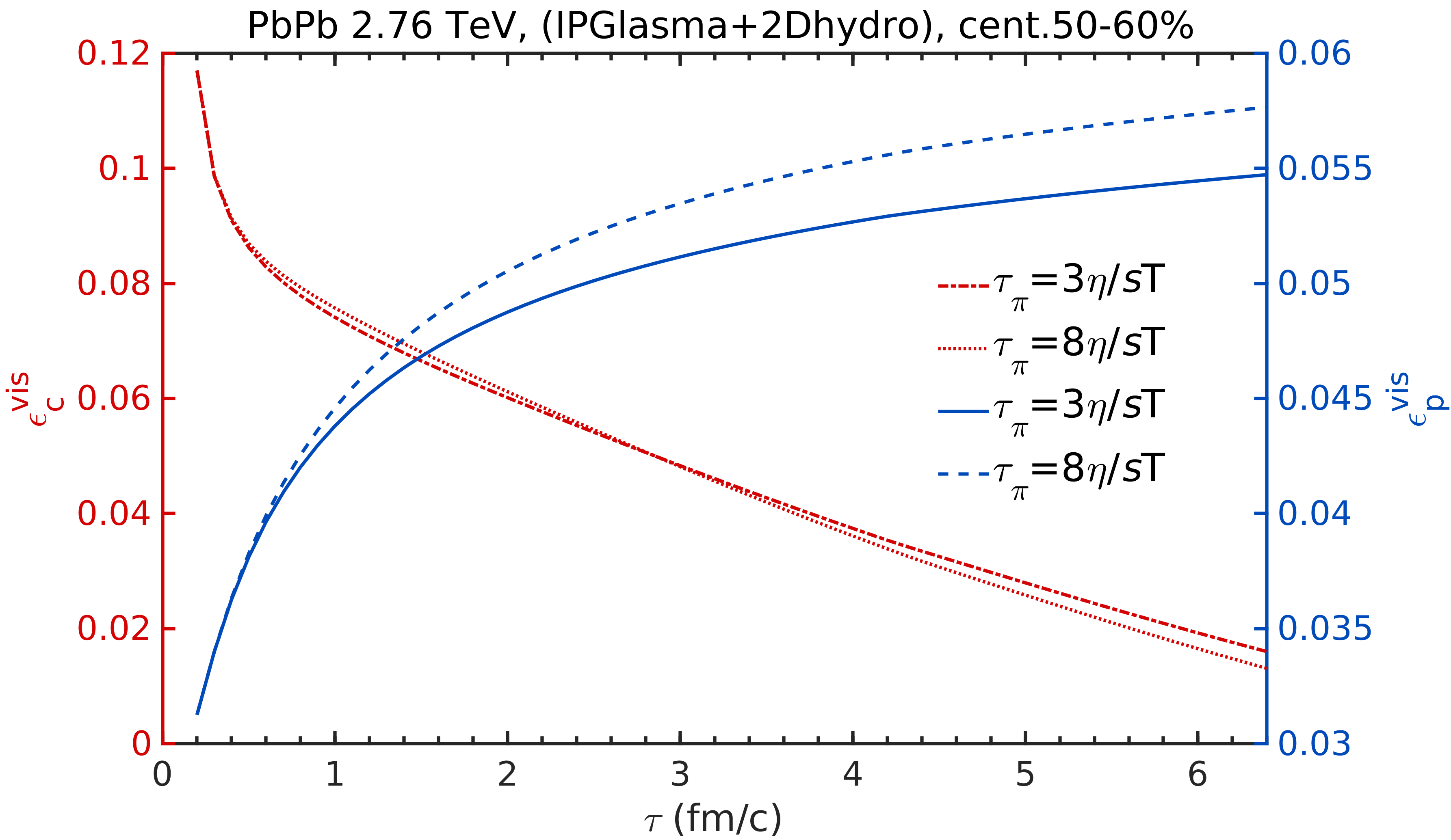}
\end{center}
\caption{Spatial eccentricity (red) and momentum space eccentricity (blue) for the viscous case for Au-Au $200$ GeV system (left) and Pb-Pb $2.76$ TeV system (right) for the two mentioned relaxation times at $50-60$\% centrality.}
\label{fig:ecc}
\end{figure*}

\begin{figure}
\includegraphics[scale=0.31]{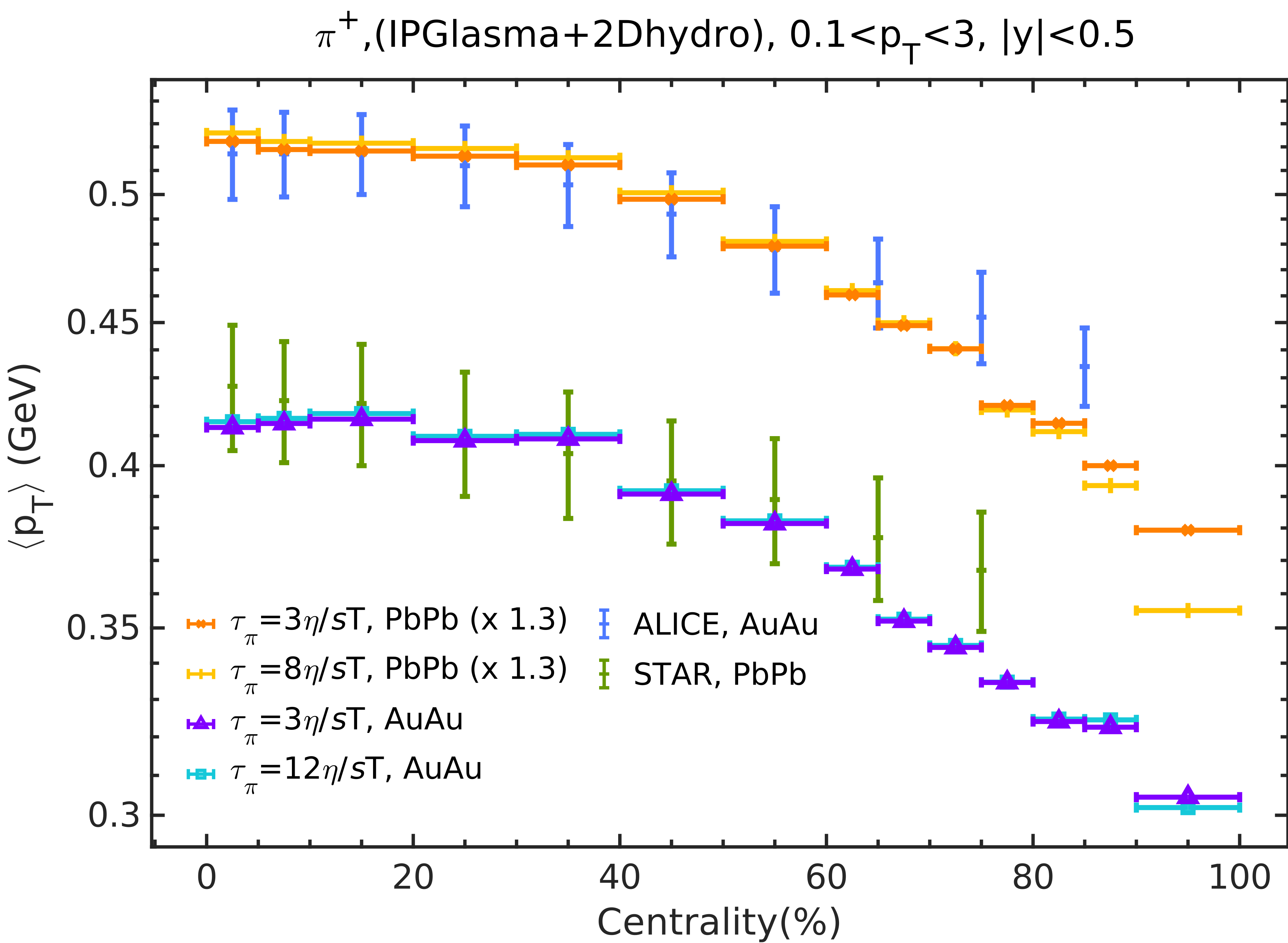}
\caption{Pion mean $p_T$ as a function of centrality for Au-Au at $200$ GeV and Pb-Pb at $2.76$ TeV. The corresponding experimental data  for PbPb from ALICE\cite{PbPb_expt_dNdy_Npart} and for AuAu from STAR\cite{EXPT_AuAu_avg_pT} have systematic errorbars.}
\label{fig:avg_pT}
\end{figure}

 \begin{figure}
\includegraphics[scale=0.7]{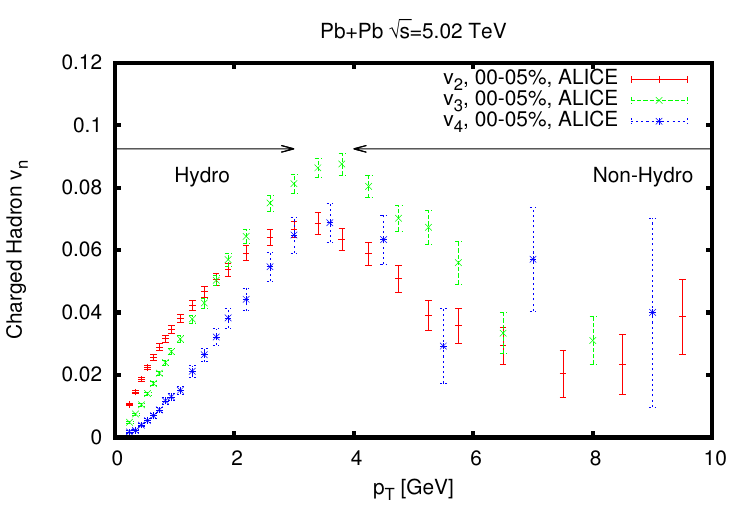}
\caption{Experimental values of flow coefficients as a function of transverse momentum. Plot taken from \cite{2017roma_app_hydro}. Phenomenological studies that make use of viscous hydrodynamics have been able to explain flow experimental data only in low $p_T$ range. Beyond $p_T\approx4$ GeV, presence of non-hydrodynamic mode has been suggested.}
\label{fig:roma_fig1}
\end{figure}

 \begin{figure*}
\includegraphics[scale=0.22]{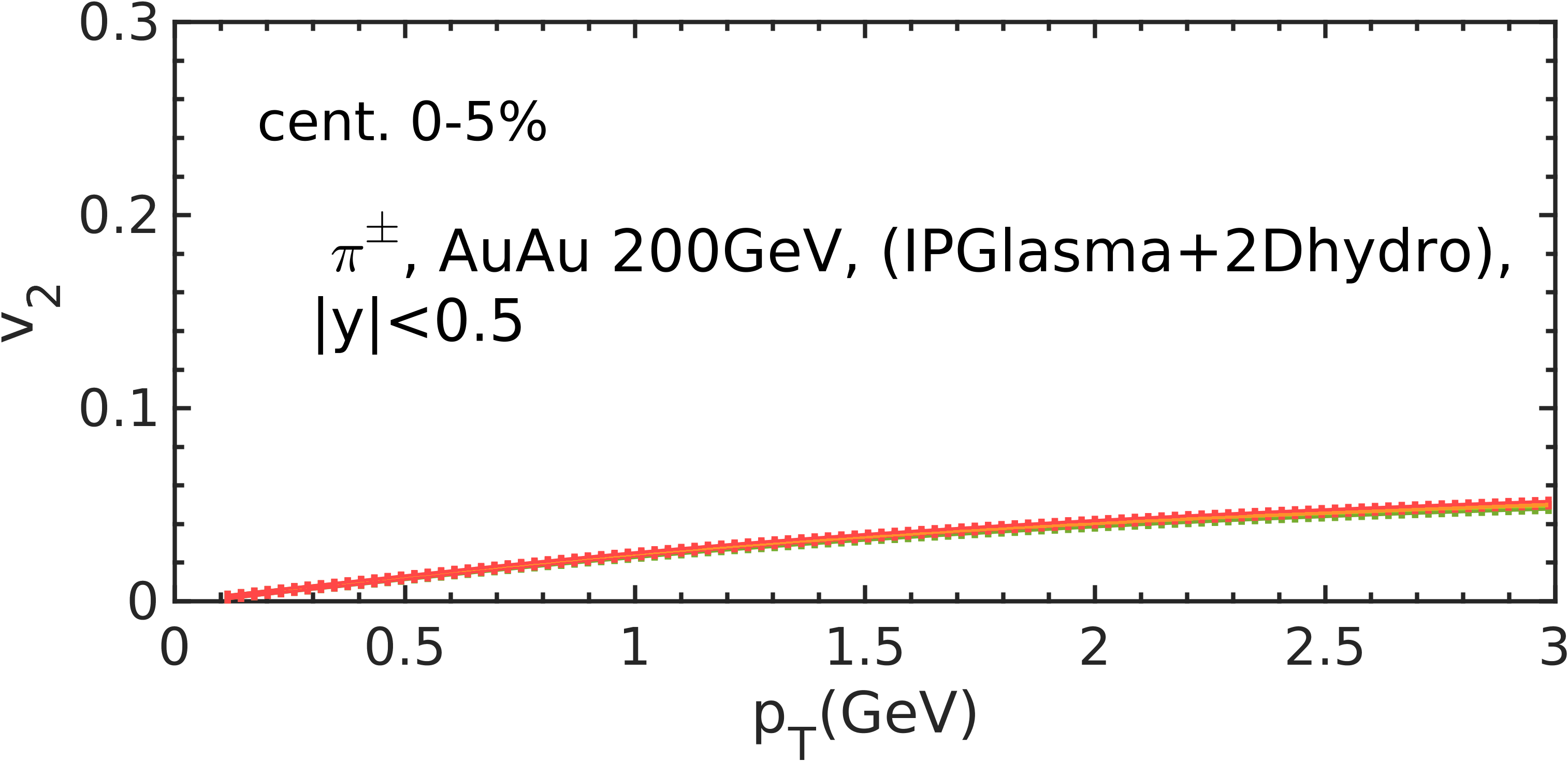}
\includegraphics[scale=0.22]{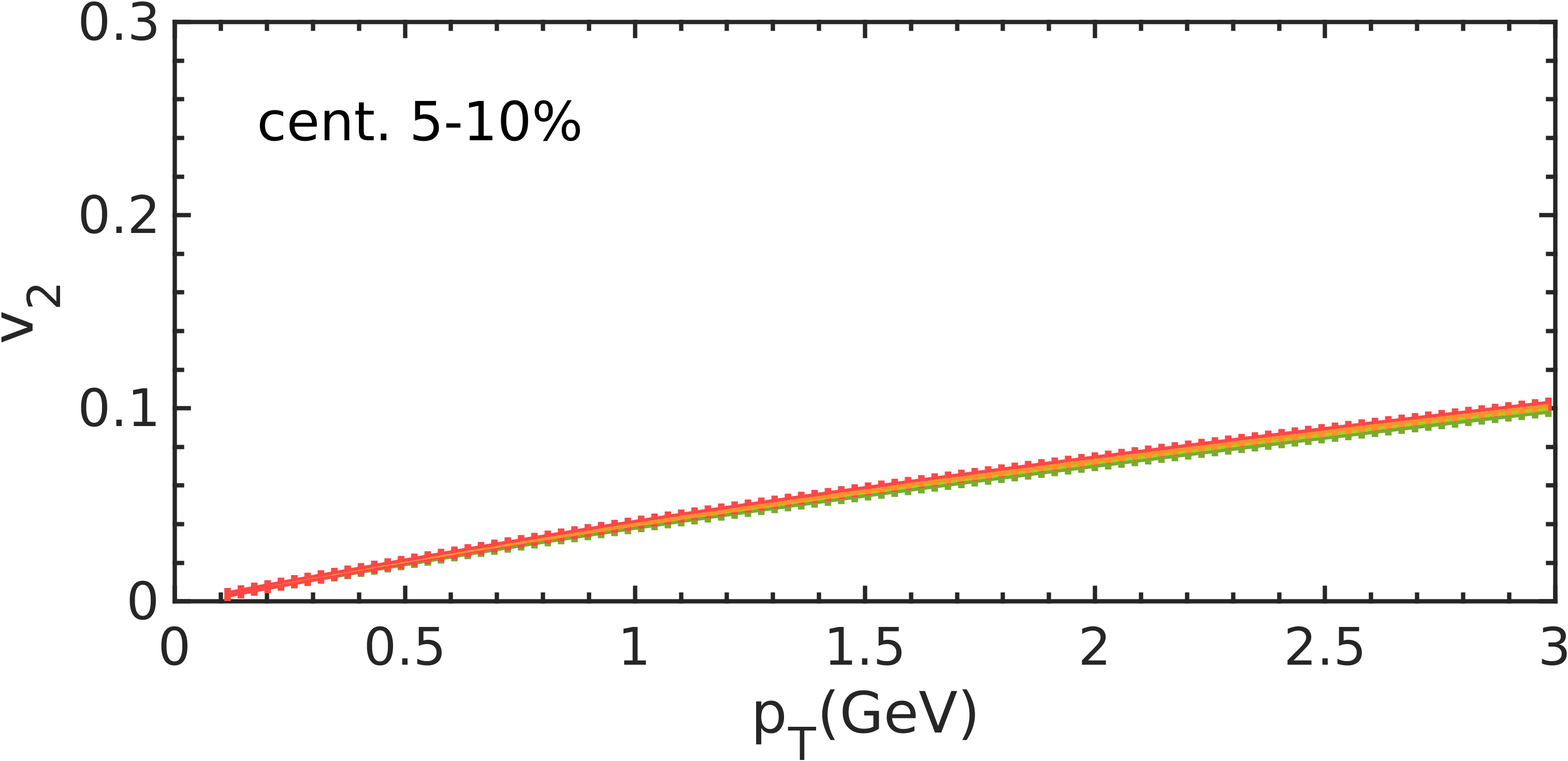}
\includegraphics[scale=0.22]{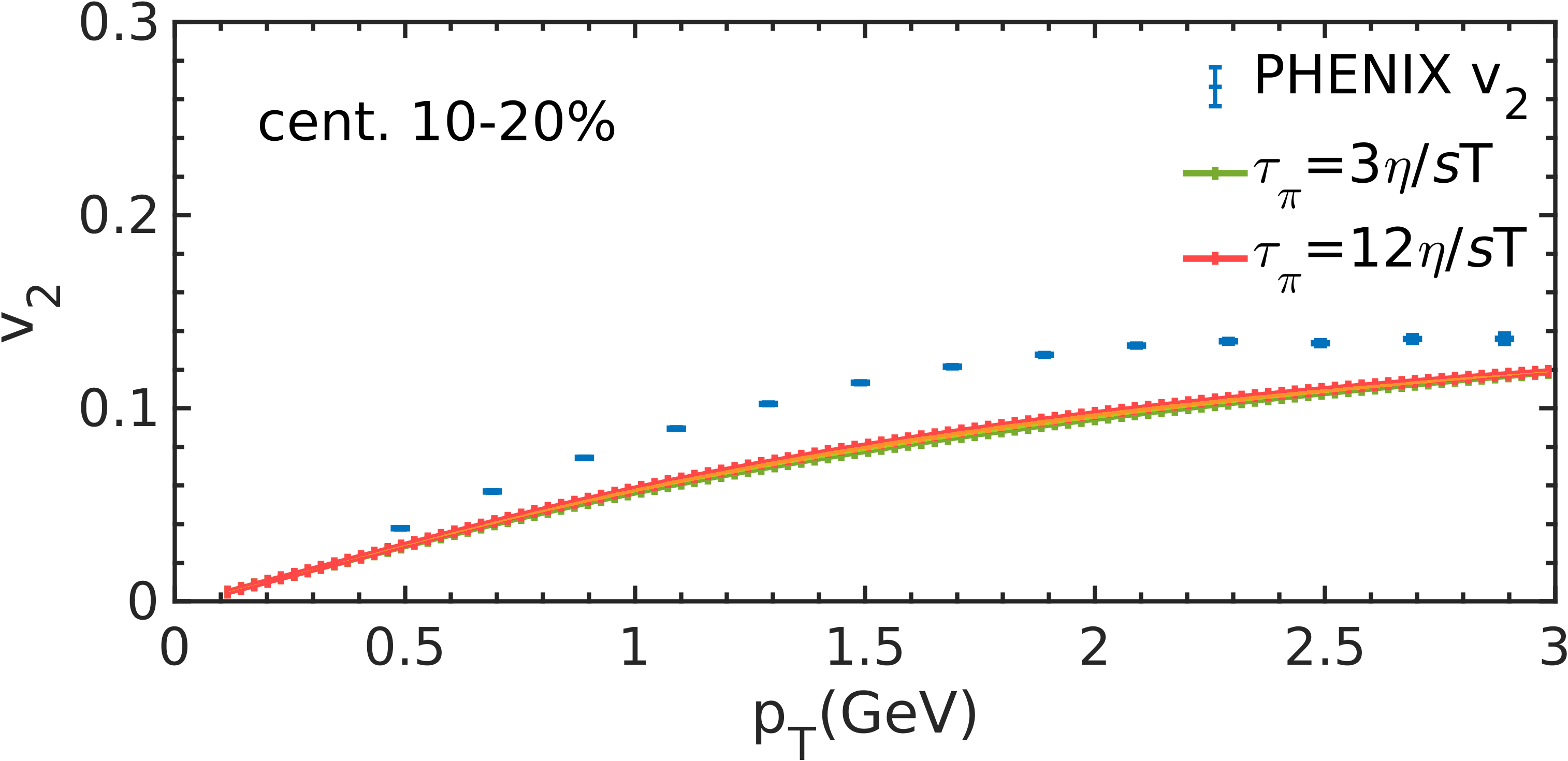}
\includegraphics[scale=0.22]{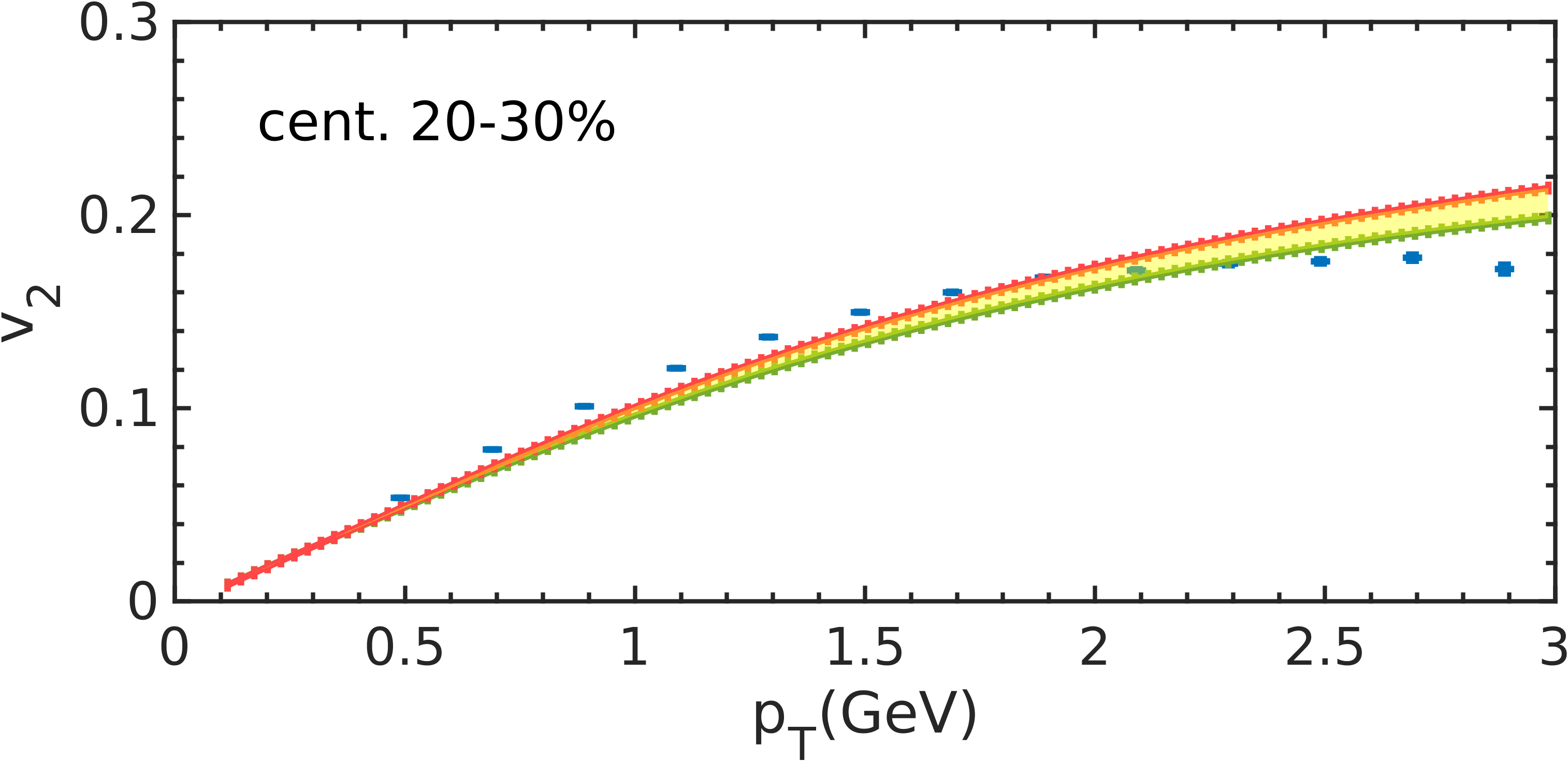}
\includegraphics[scale=0.22]{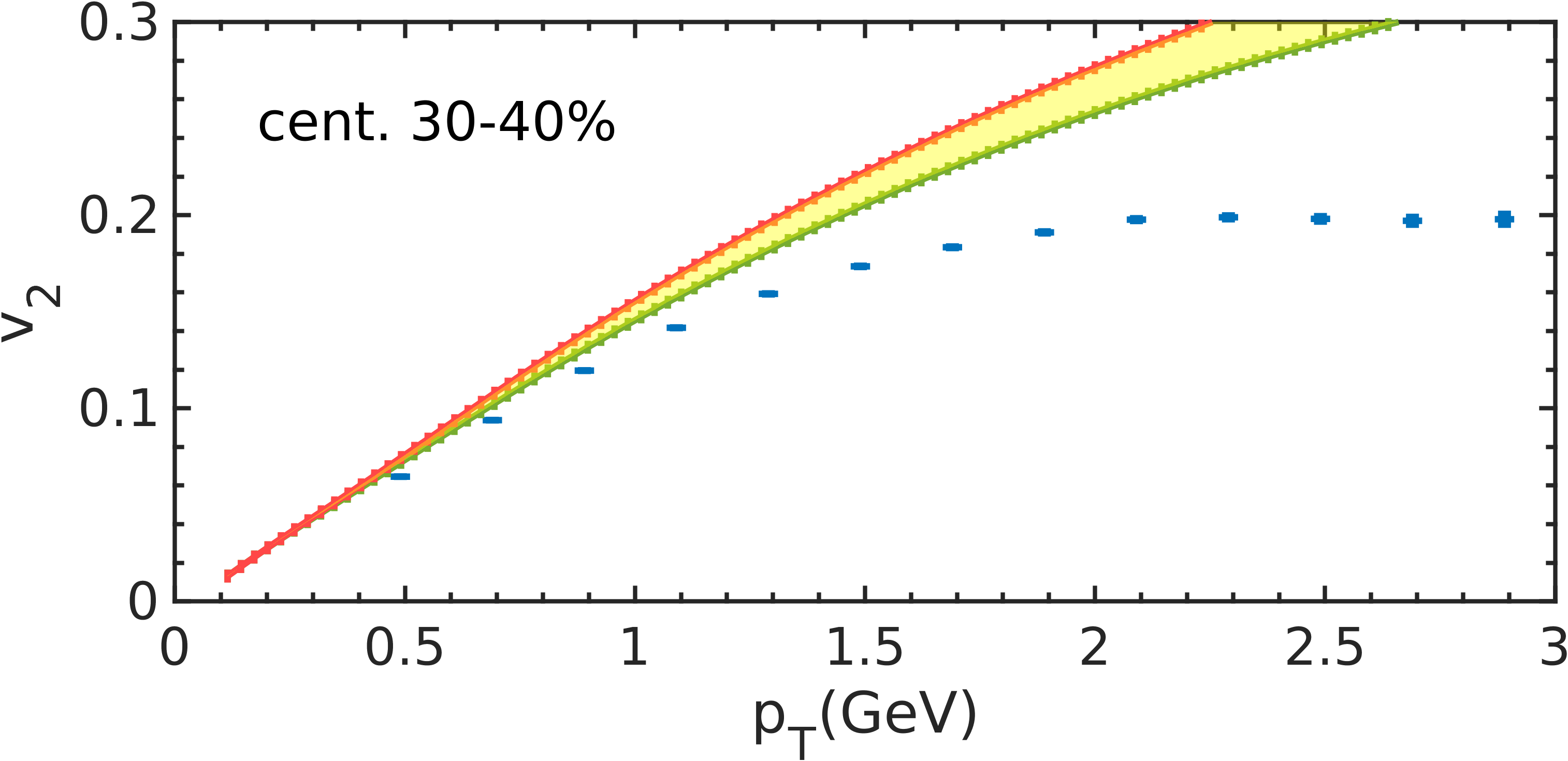}
\includegraphics[scale=0.22]{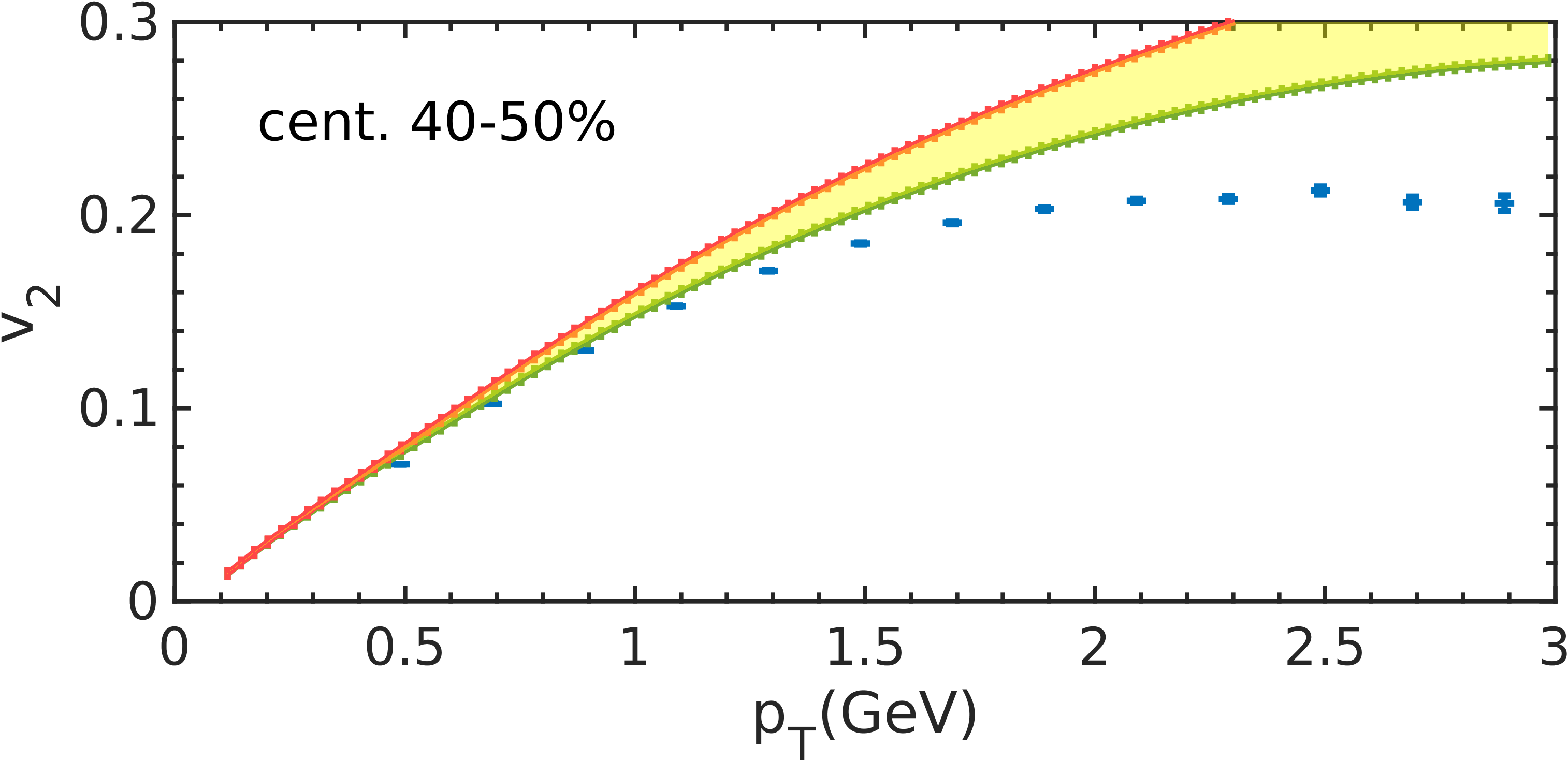}
\includegraphics[scale=0.22]{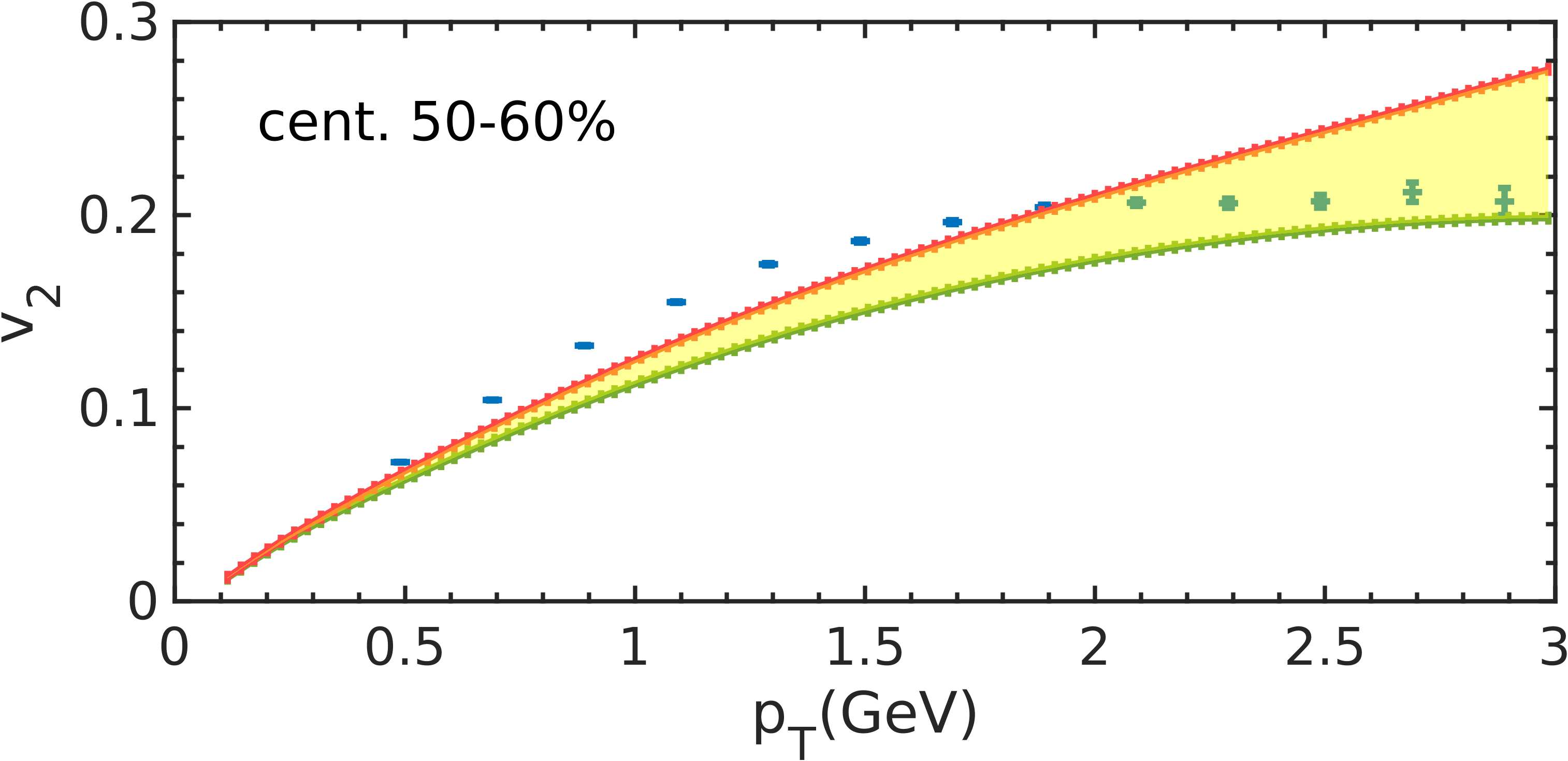}
\includegraphics[scale=0.22]{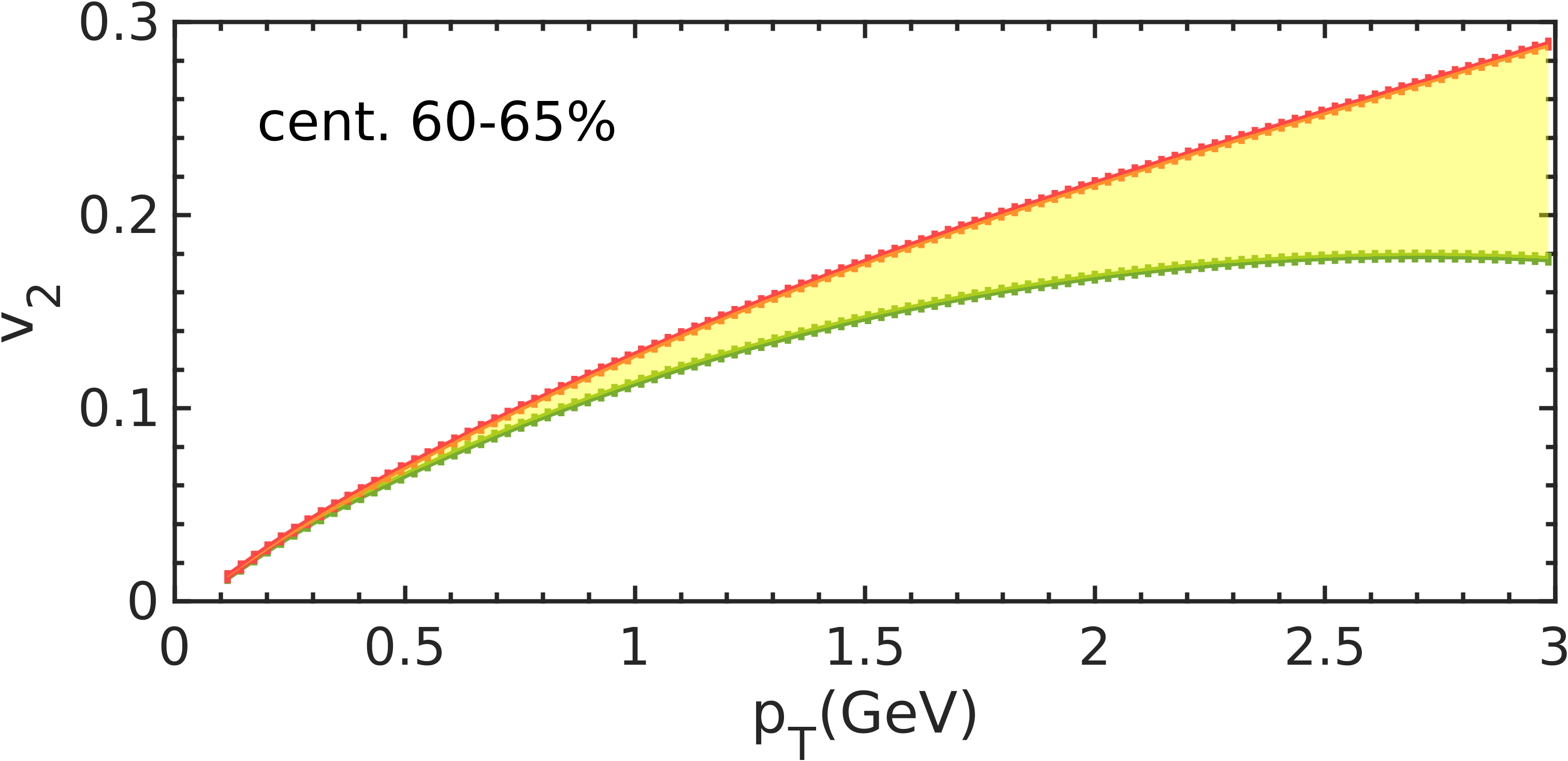}
\includegraphics[scale=0.22]{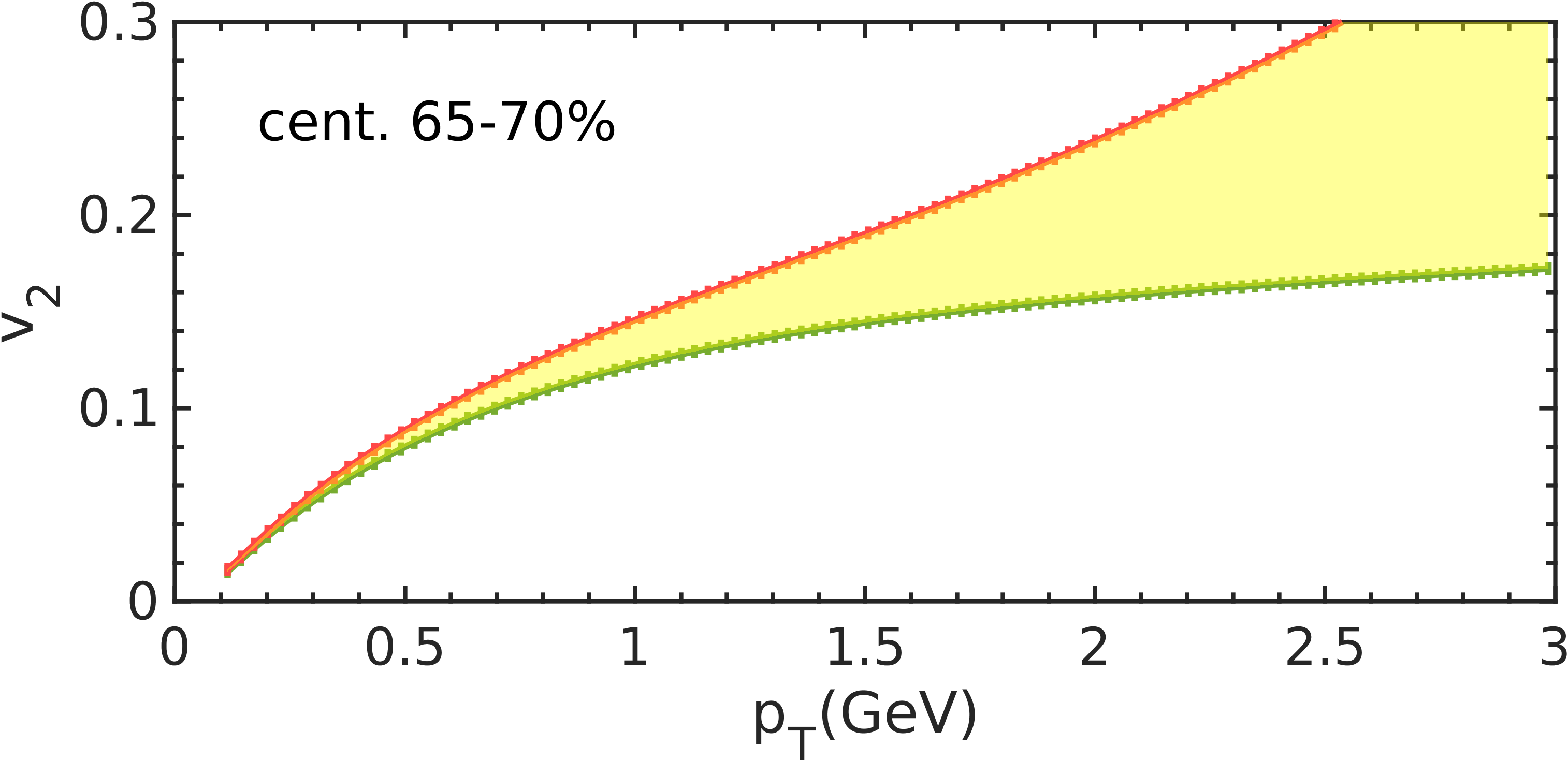}
\includegraphics[scale=0.22]{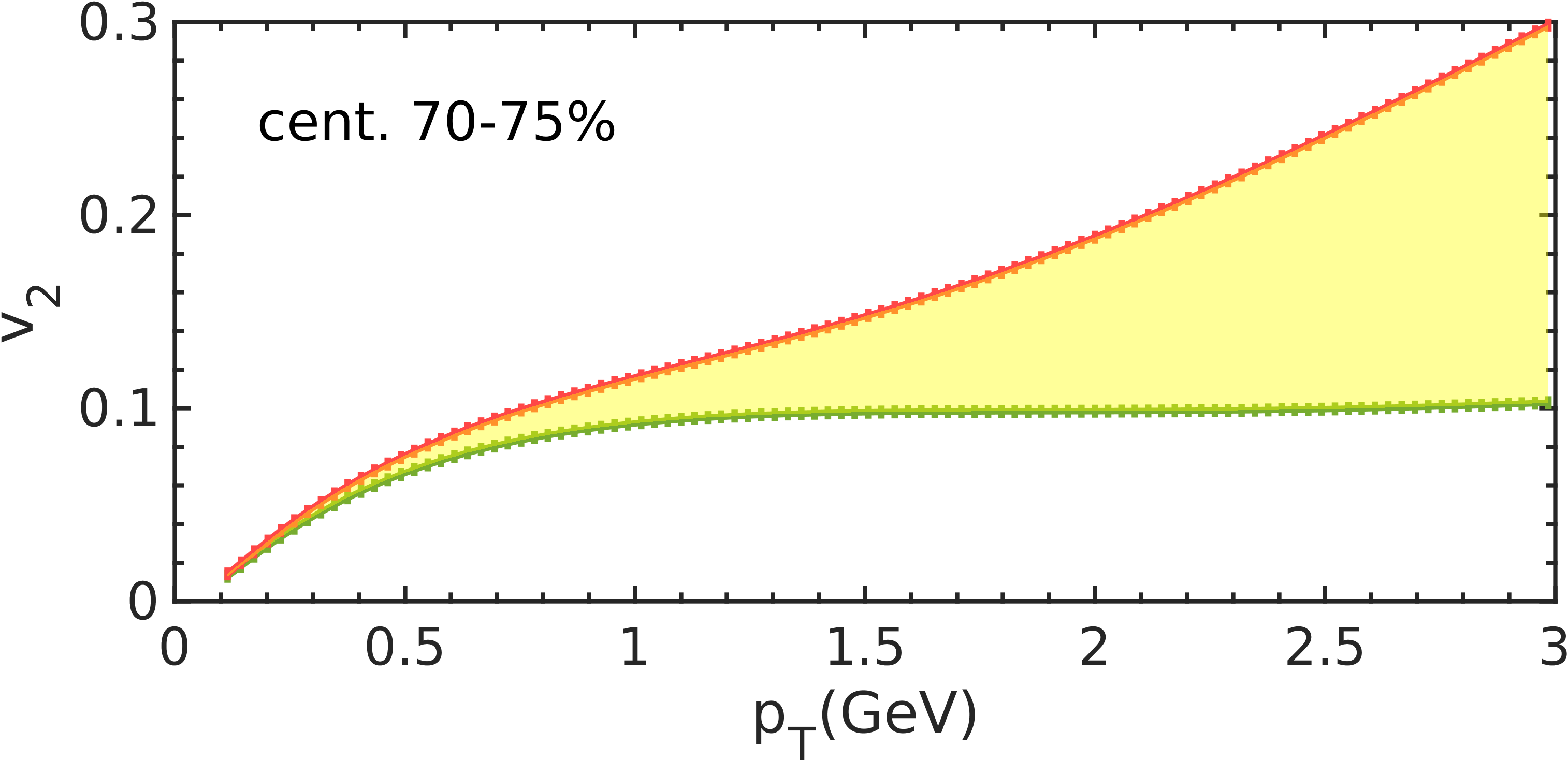}
\includegraphics[scale=0.22]{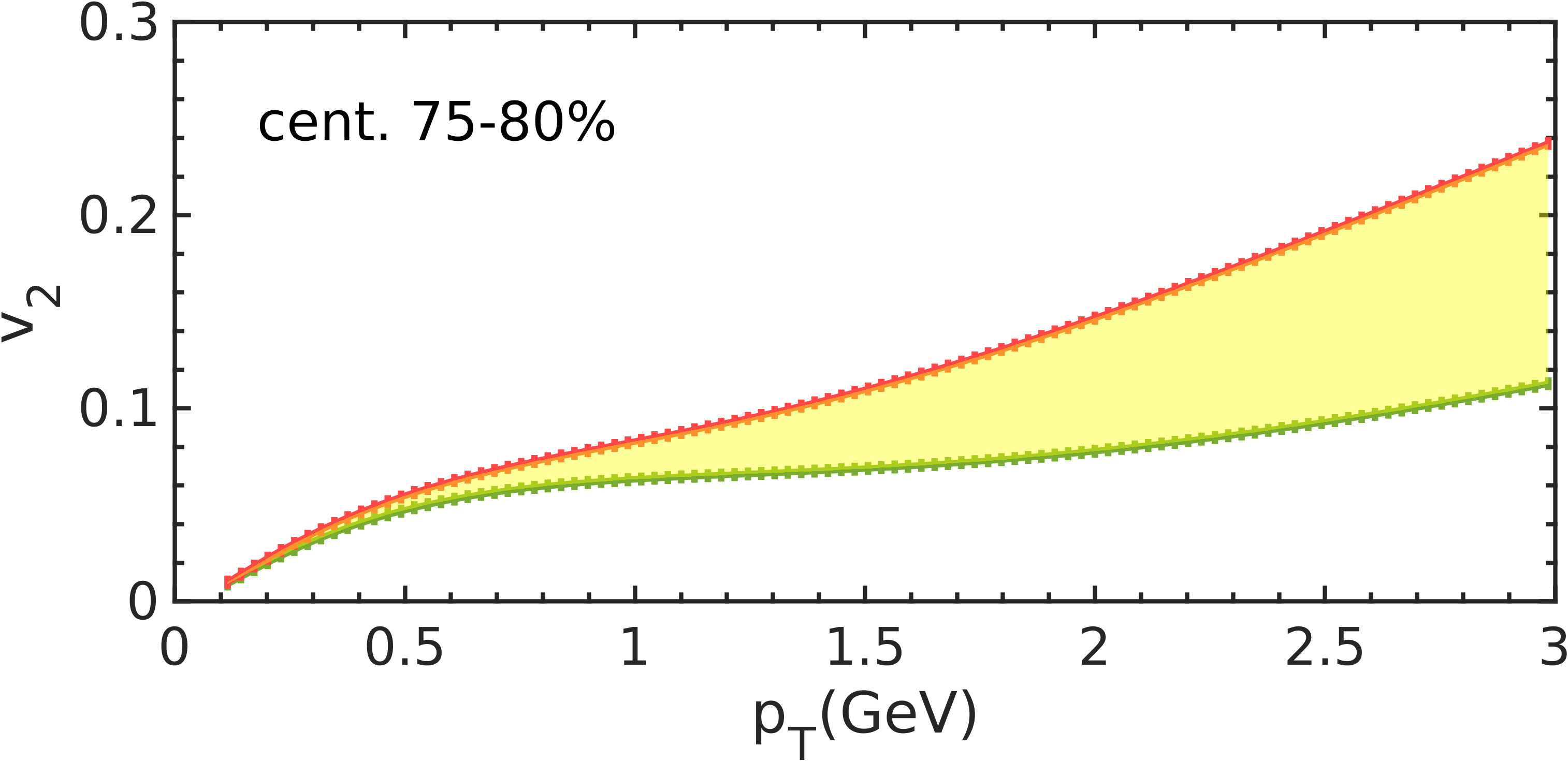}
\includegraphics[scale=0.22]{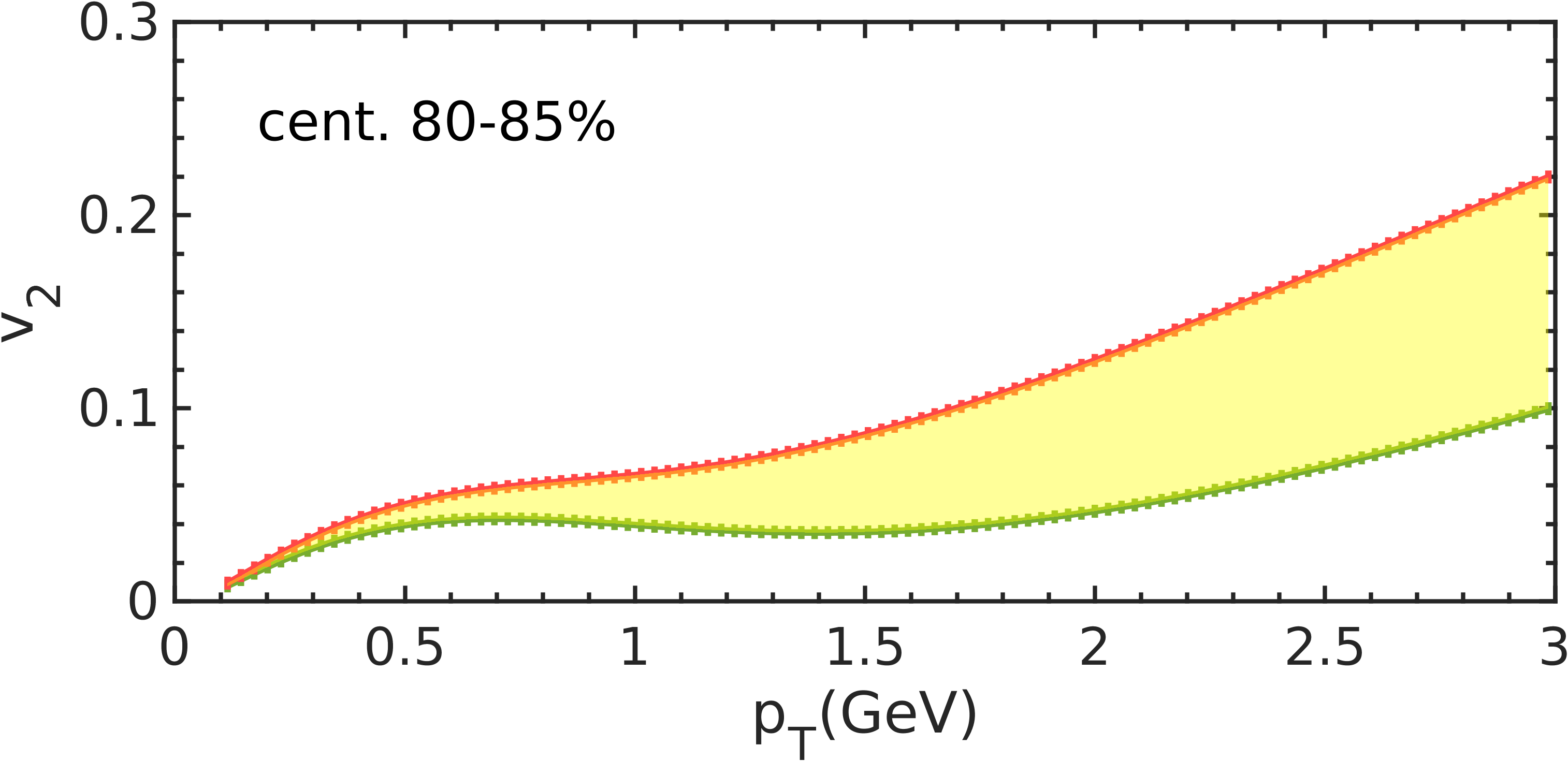}
\includegraphics[scale=0.22]{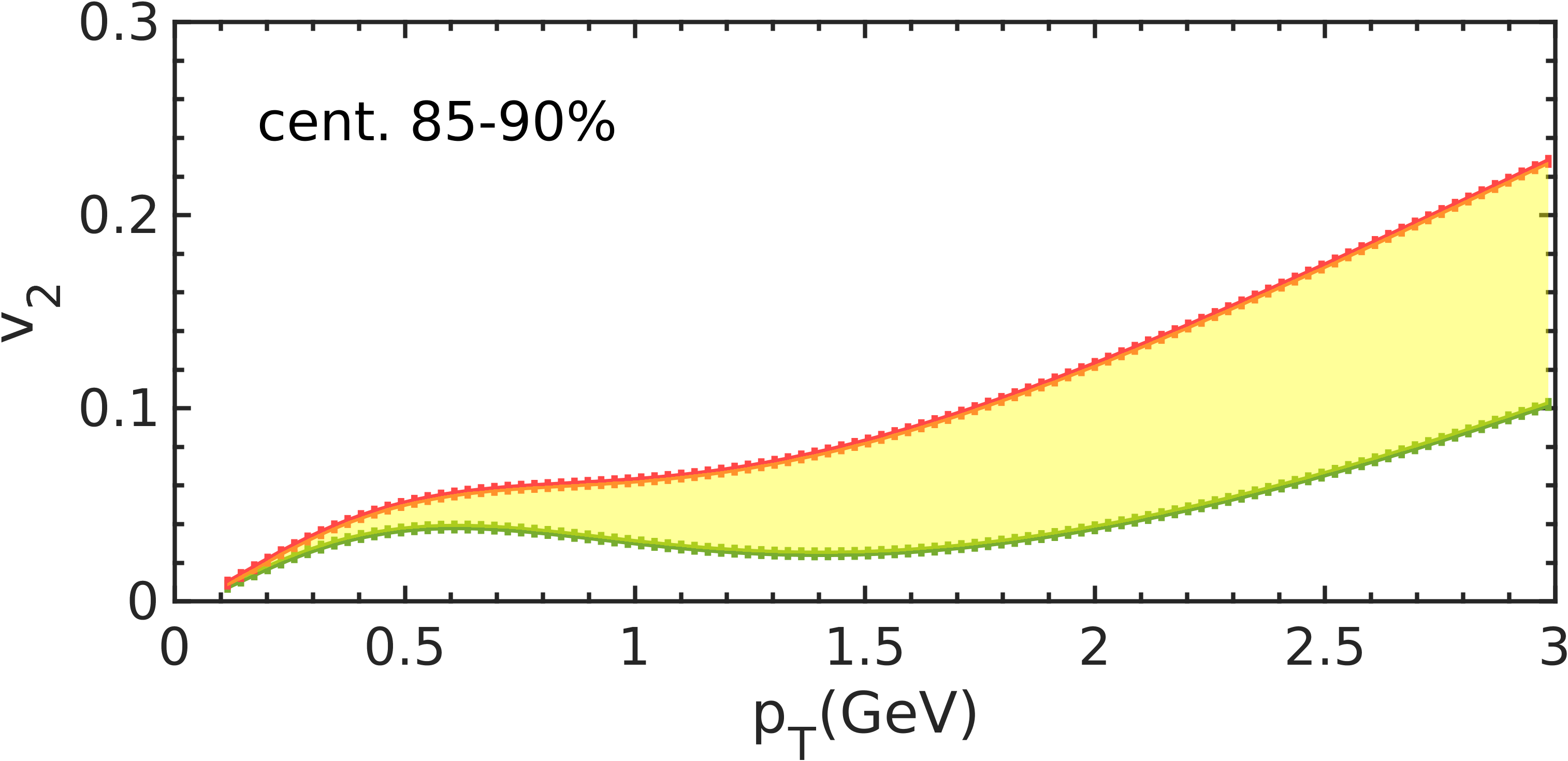}
\includegraphics[scale=0.22]{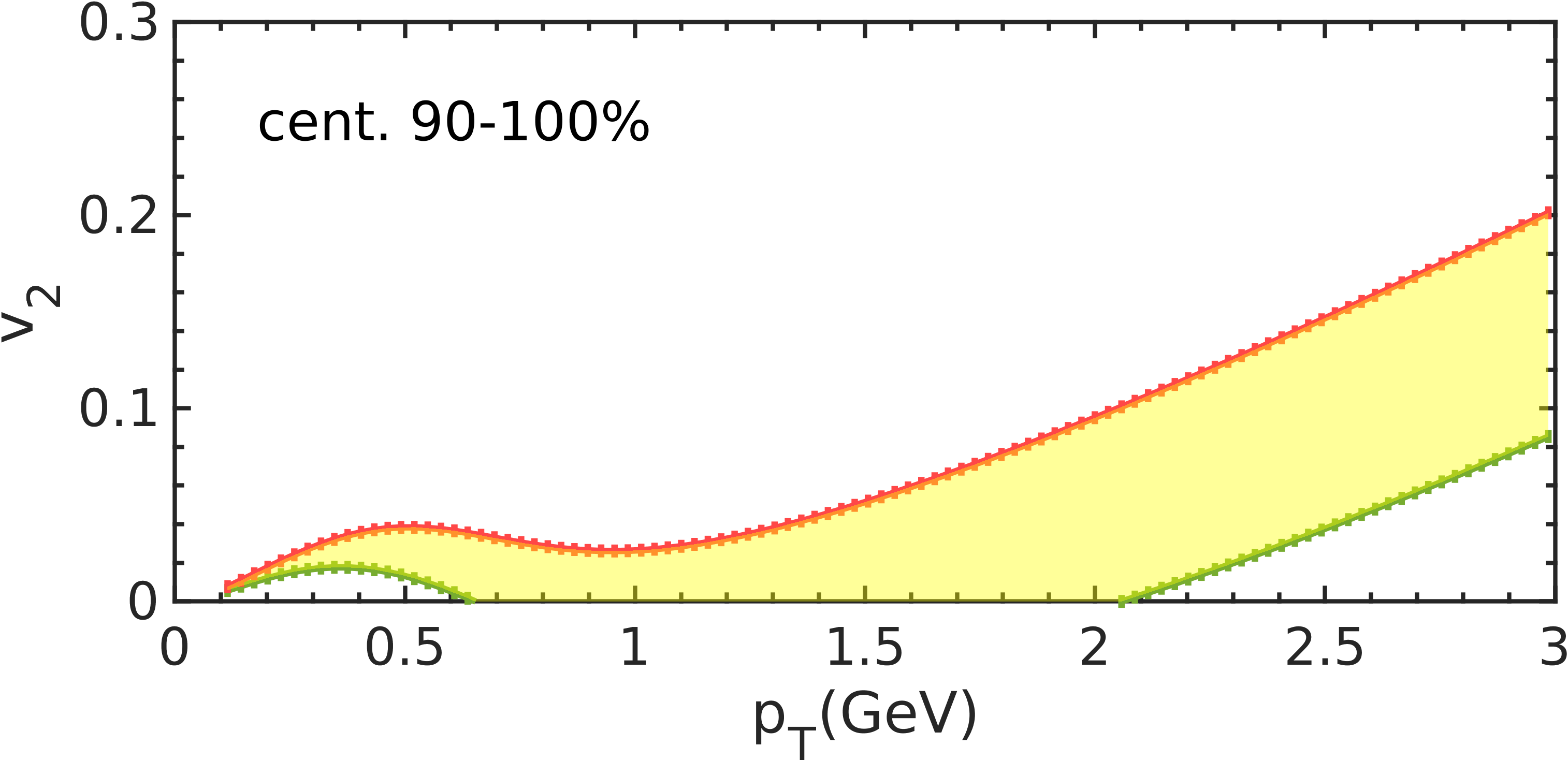}
\caption{Pion($\pi^+$) elliptic flow coefficient($v_2$) as a function of transverse momentum($p_T$) for $14$ centrality classes for Au-Au $200$ GeV system obtained with (IPGlasma+2Dhydro) set up along with experimentally measured elliptic flw (blue) from PHENIX \cite{EXPT_AuAu_v2_pT} for the relaxation/non-hydrodynamic mode decay time, $\tau_\pi =3\eta/sT$ (green) and $12\eta/sT$(red). The shaded area (yellow) highlights the difference in flow due to variation in relaxation time.}
\label{fig:pT_v2_AuAu}
\end{figure*}

 \begin{figure*}
\includegraphics[scale=0.22]{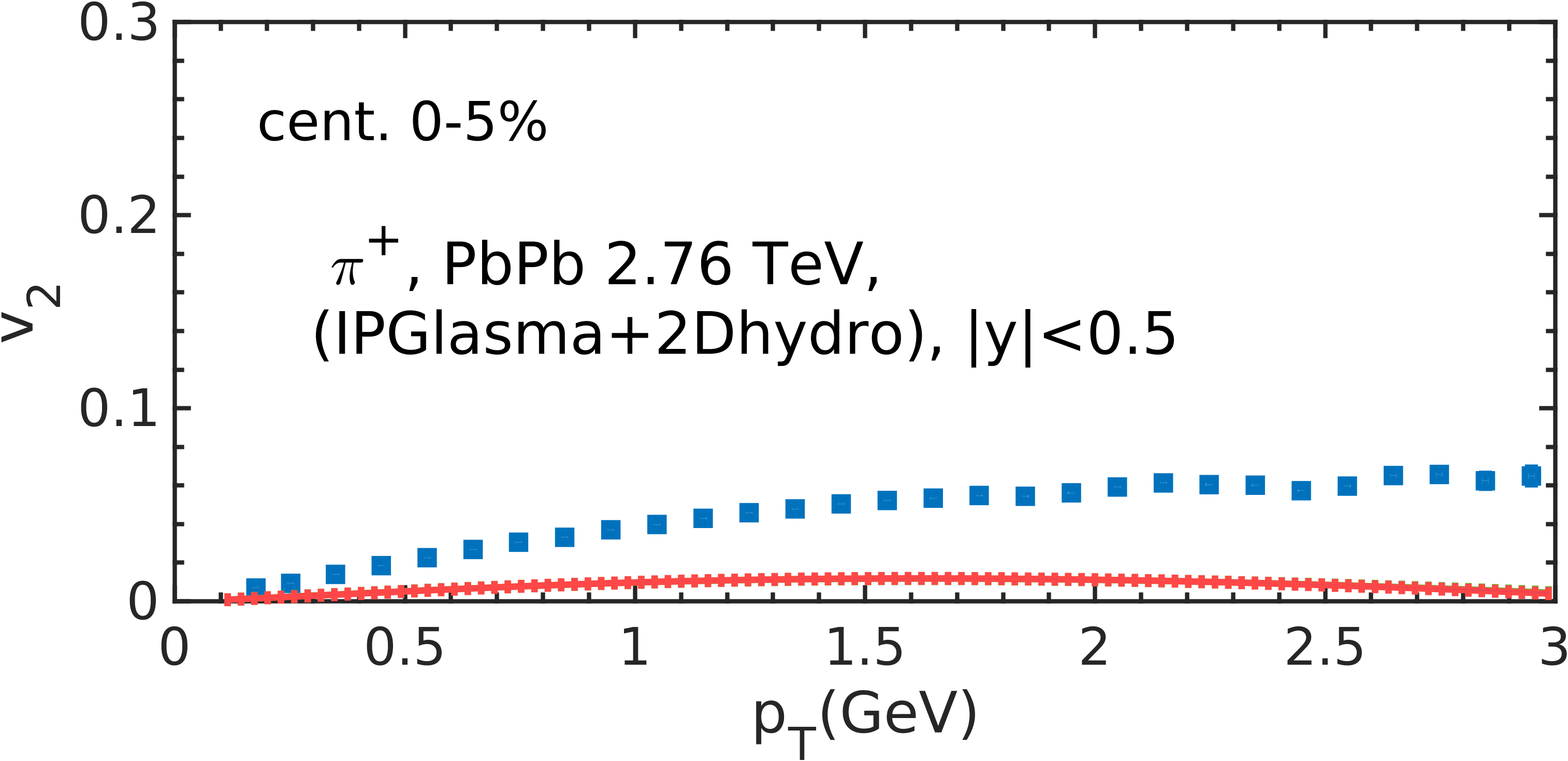}
\includegraphics[scale=0.22]{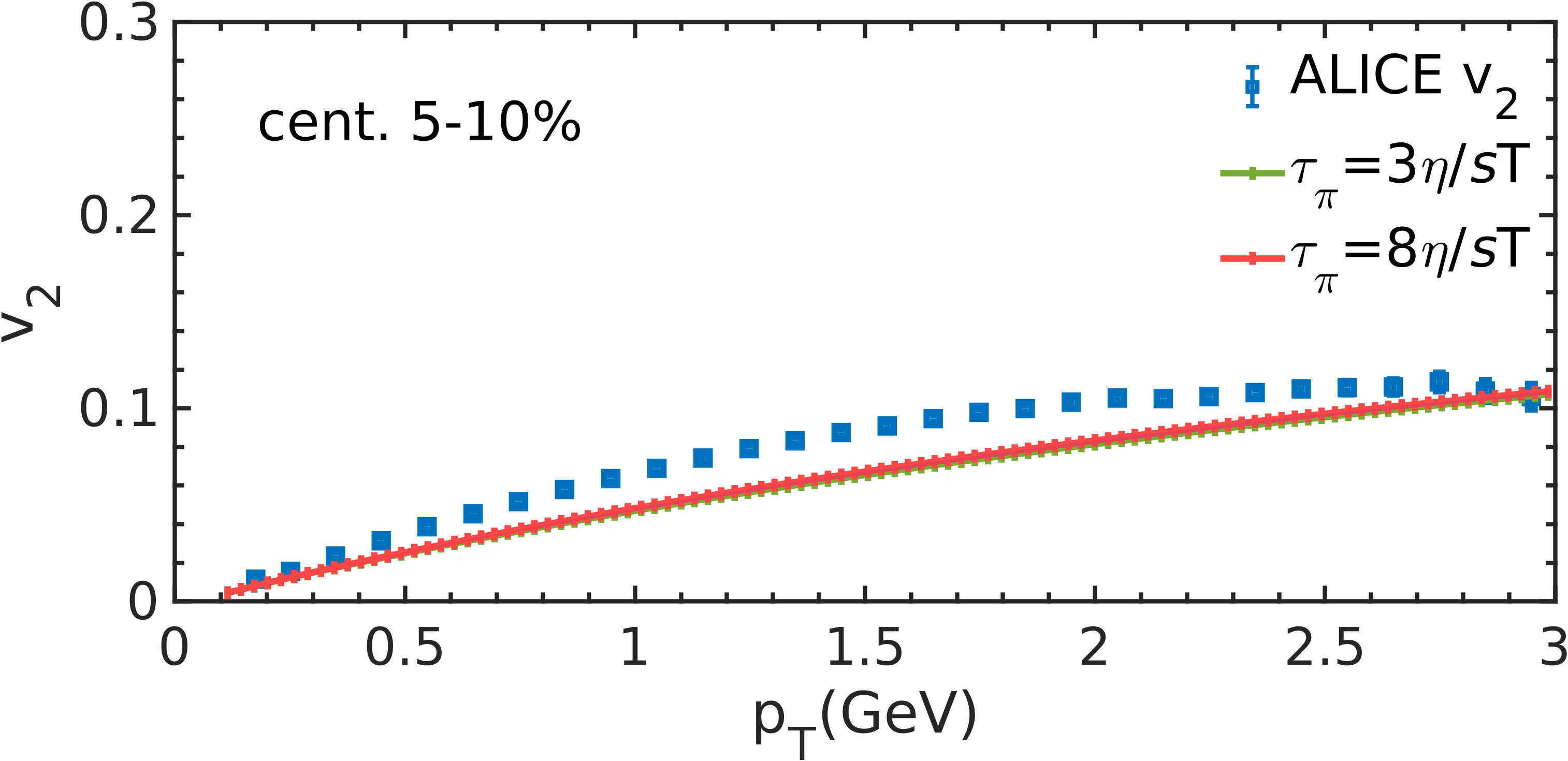}
\includegraphics[scale=0.22]{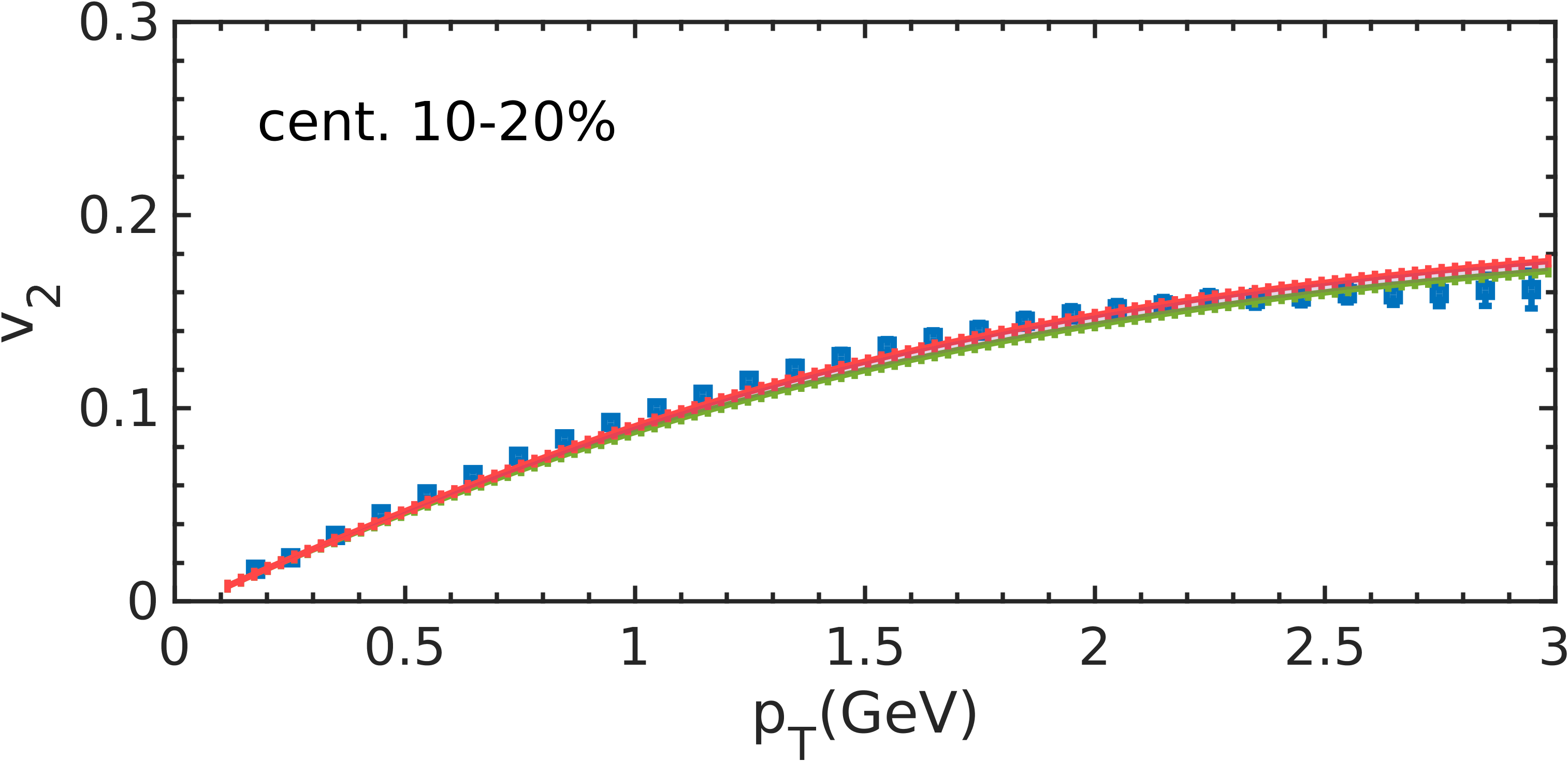}
\includegraphics[scale=0.22]{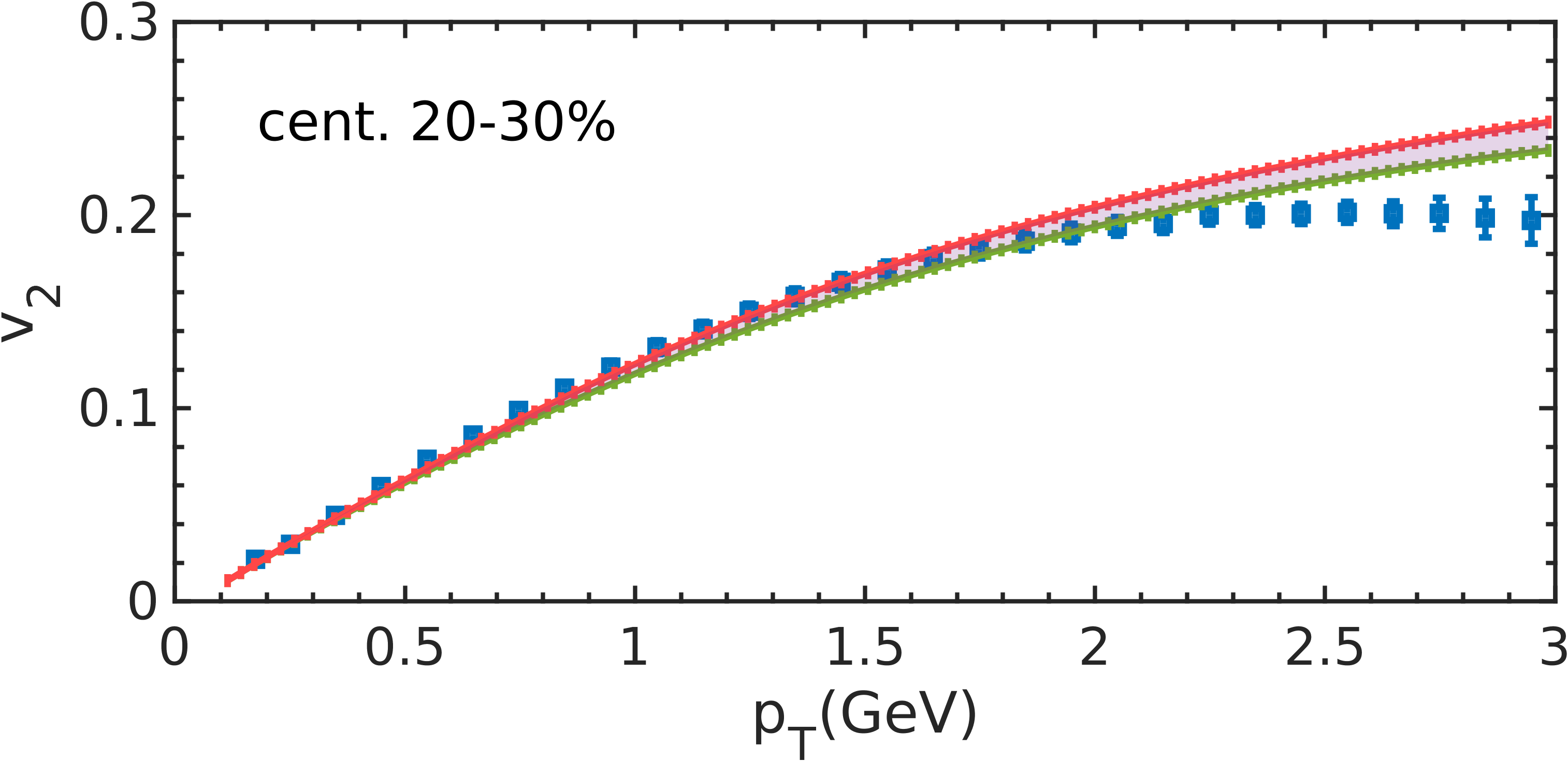}
\includegraphics[scale=0.22]{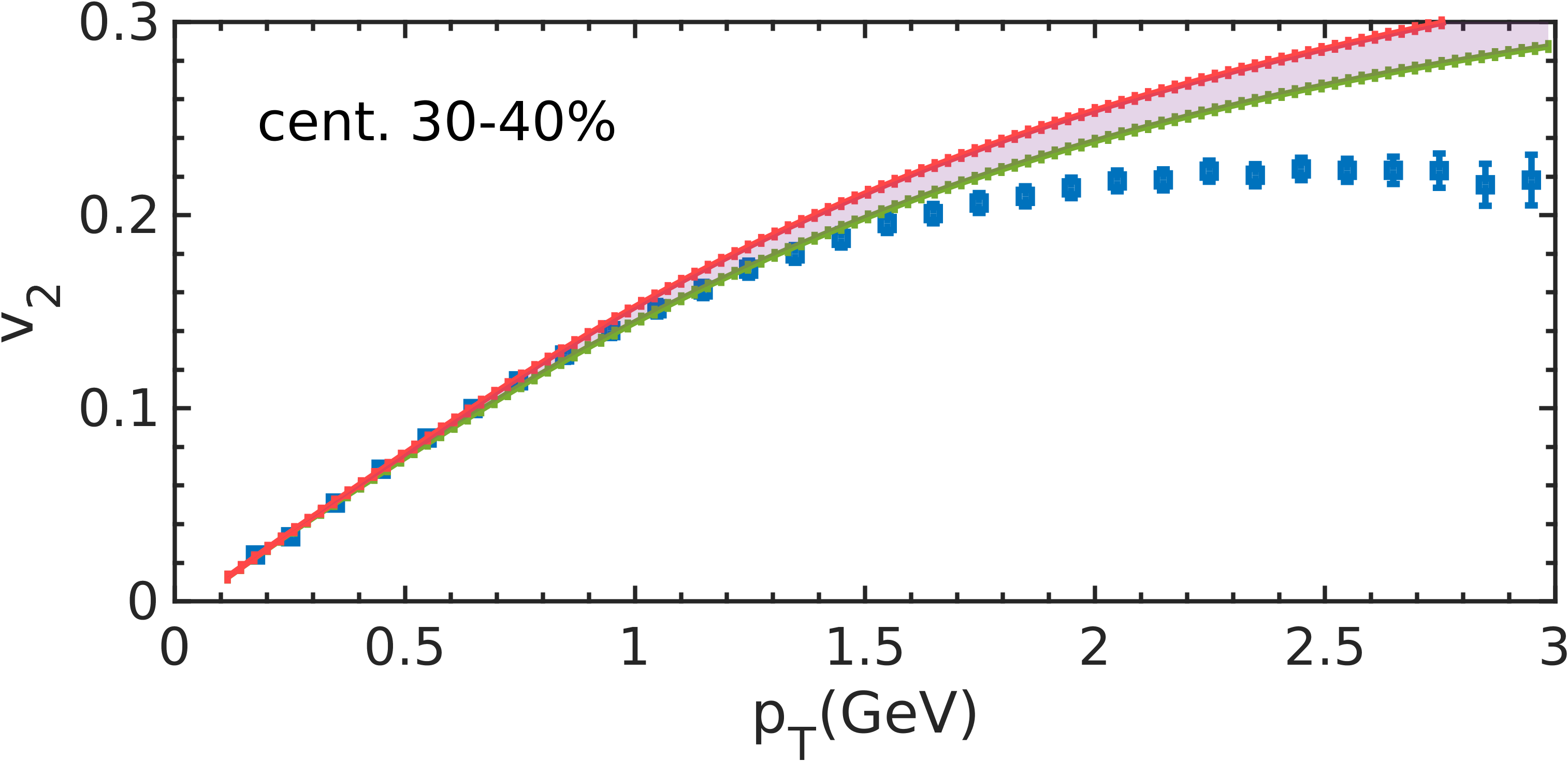}
\includegraphics[scale=0.22]{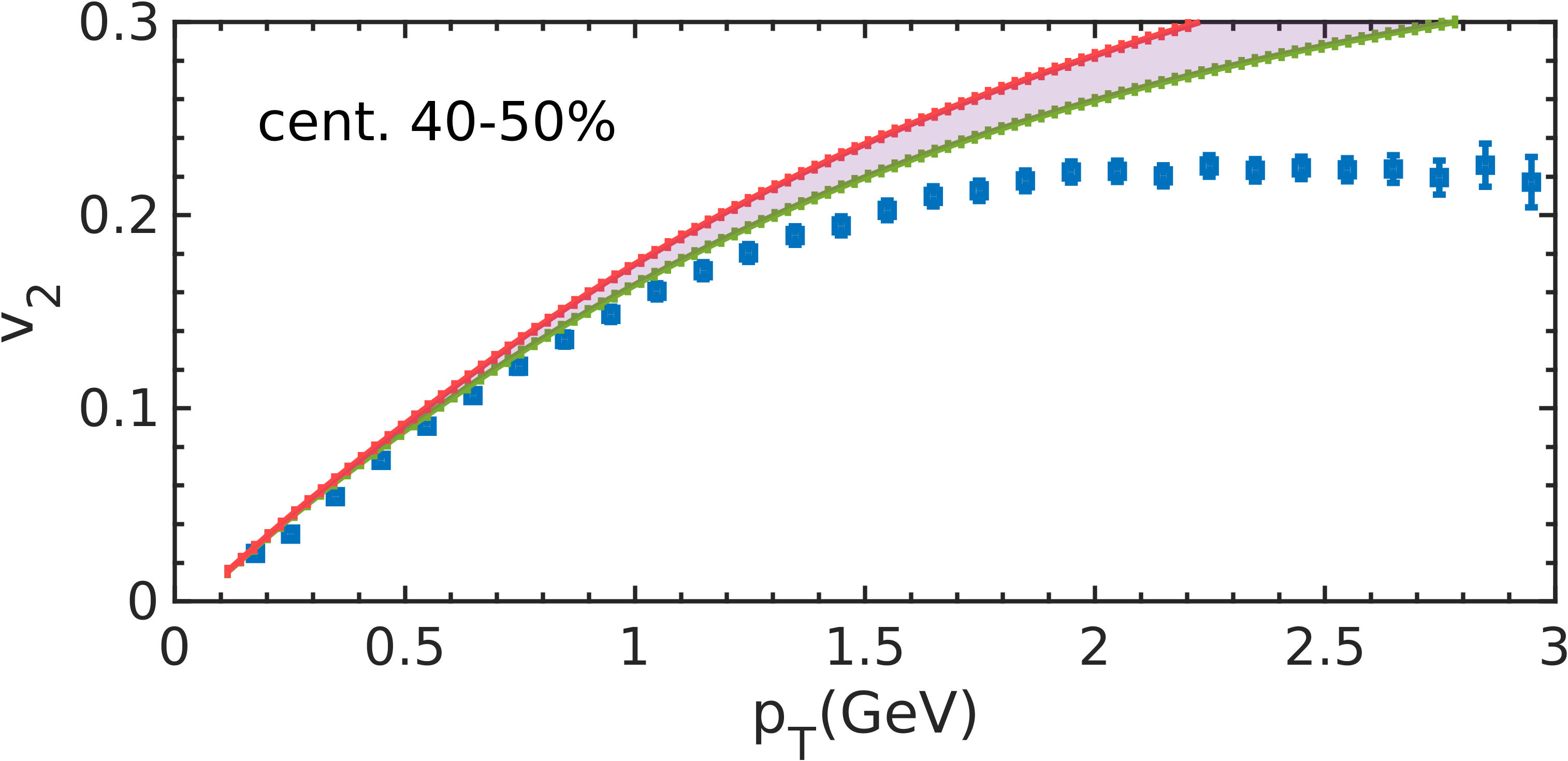}
\includegraphics[scale=0.22]{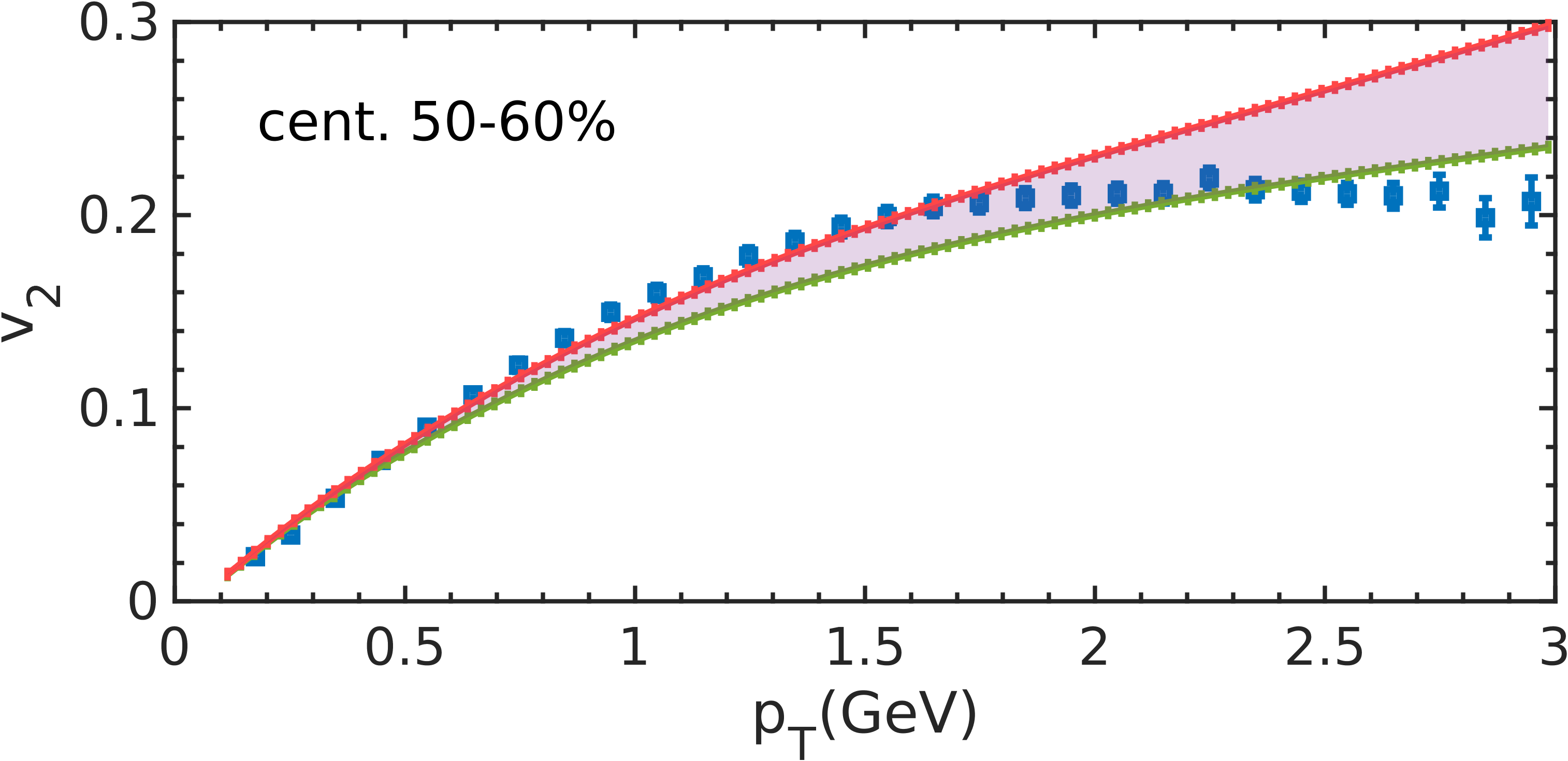}
\includegraphics[scale=0.22]{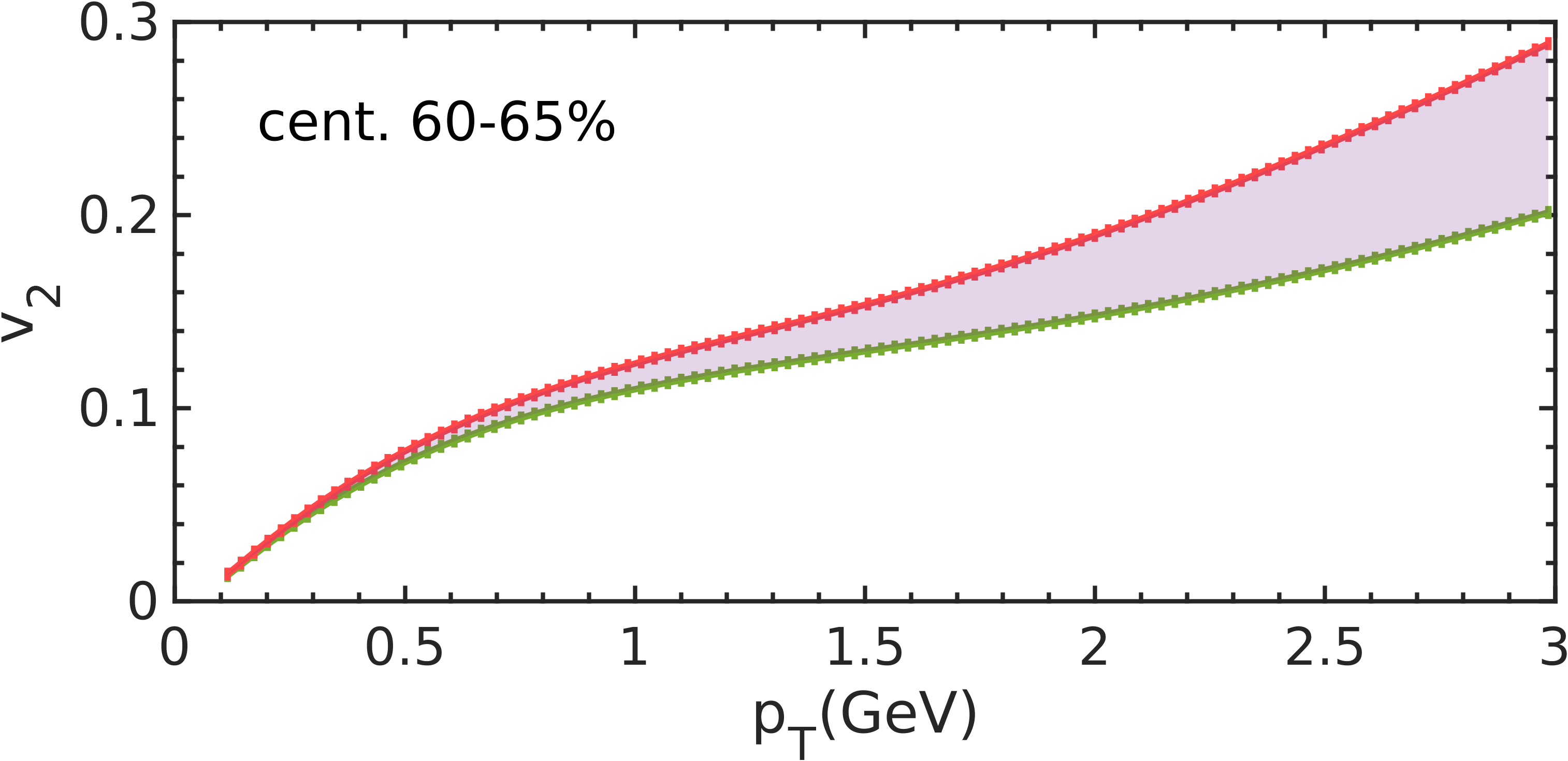}
\includegraphics[scale=0.22]{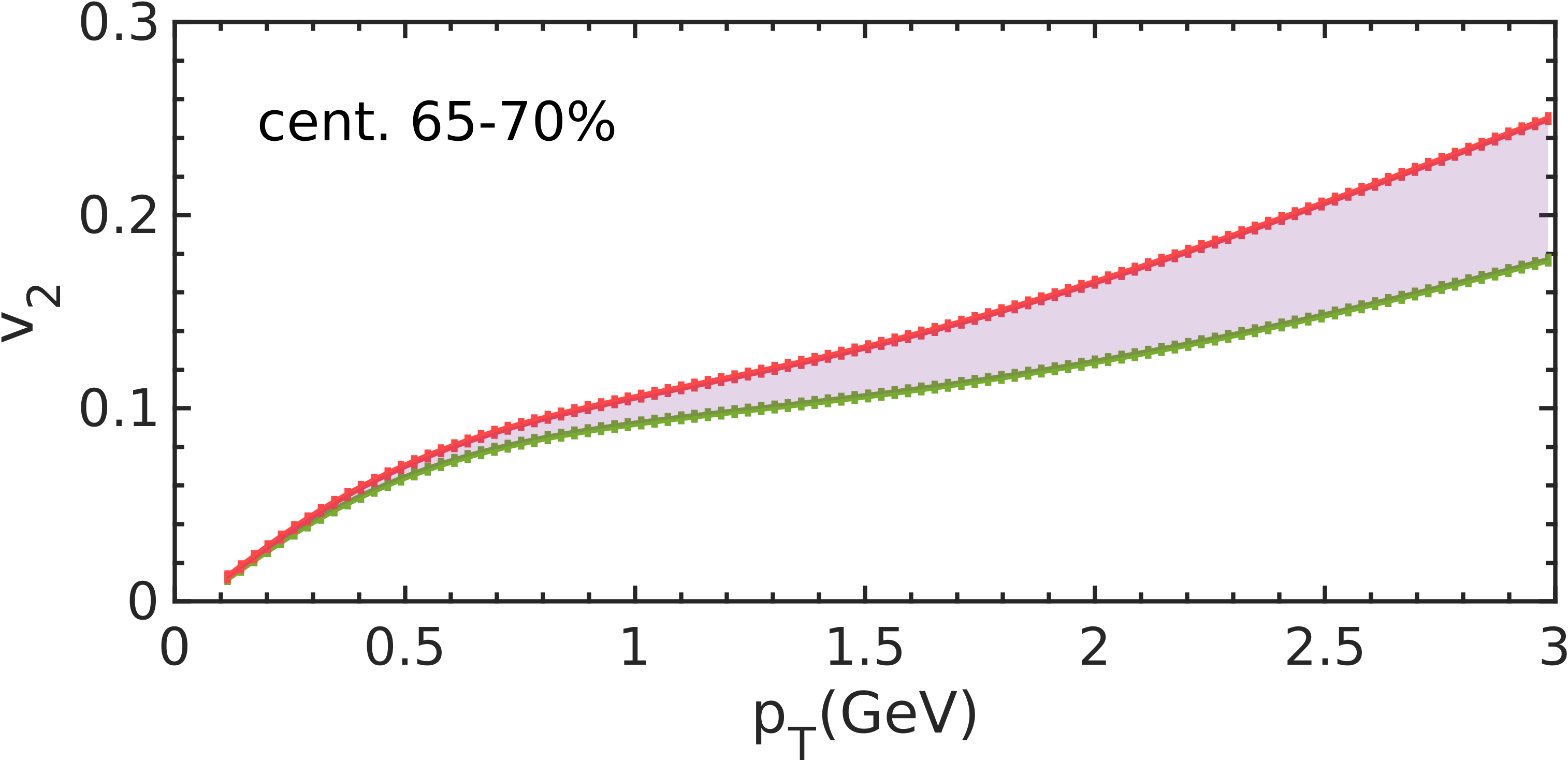}
\includegraphics[scale=0.22]{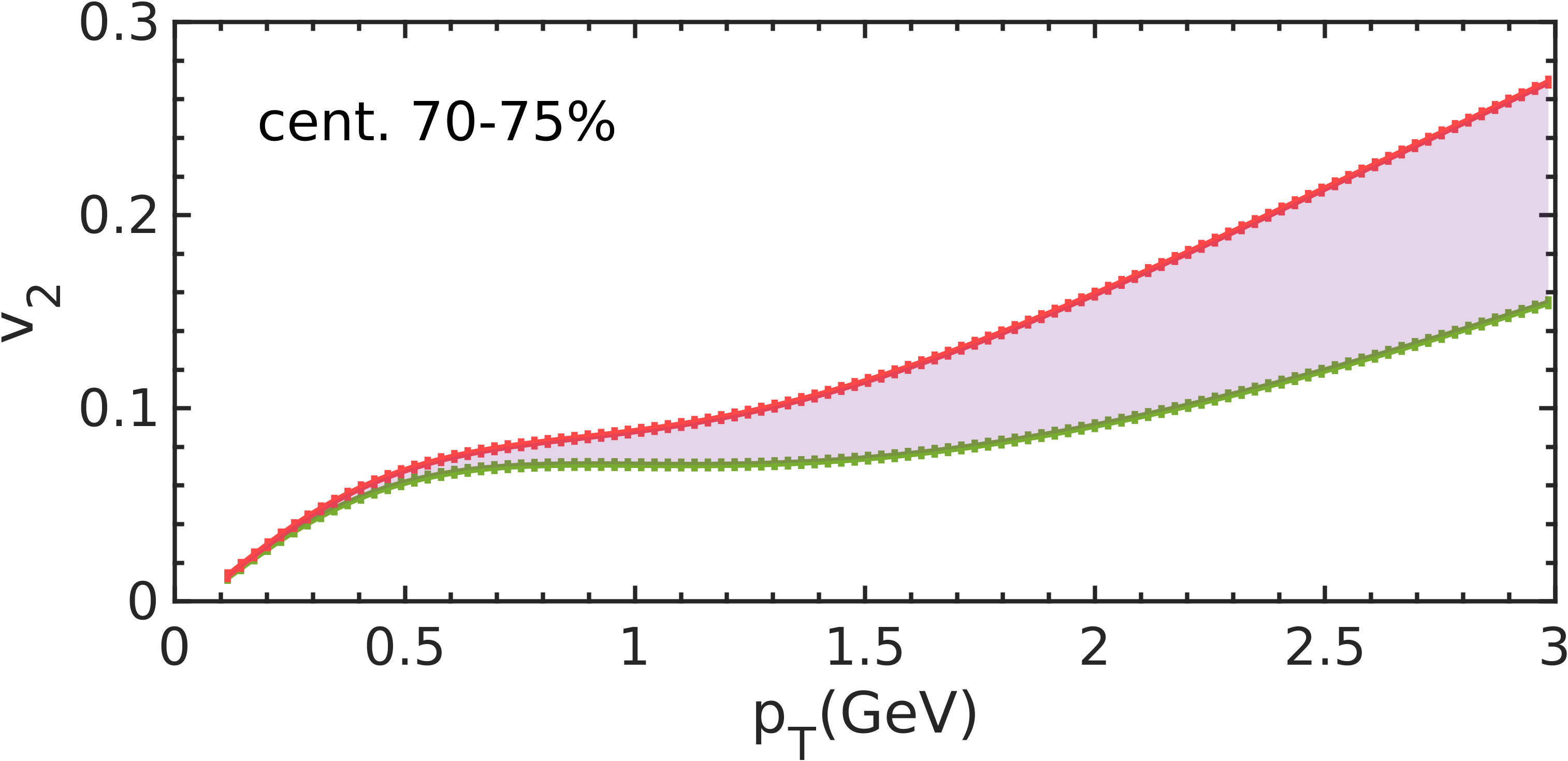}
\includegraphics[scale=0.22]{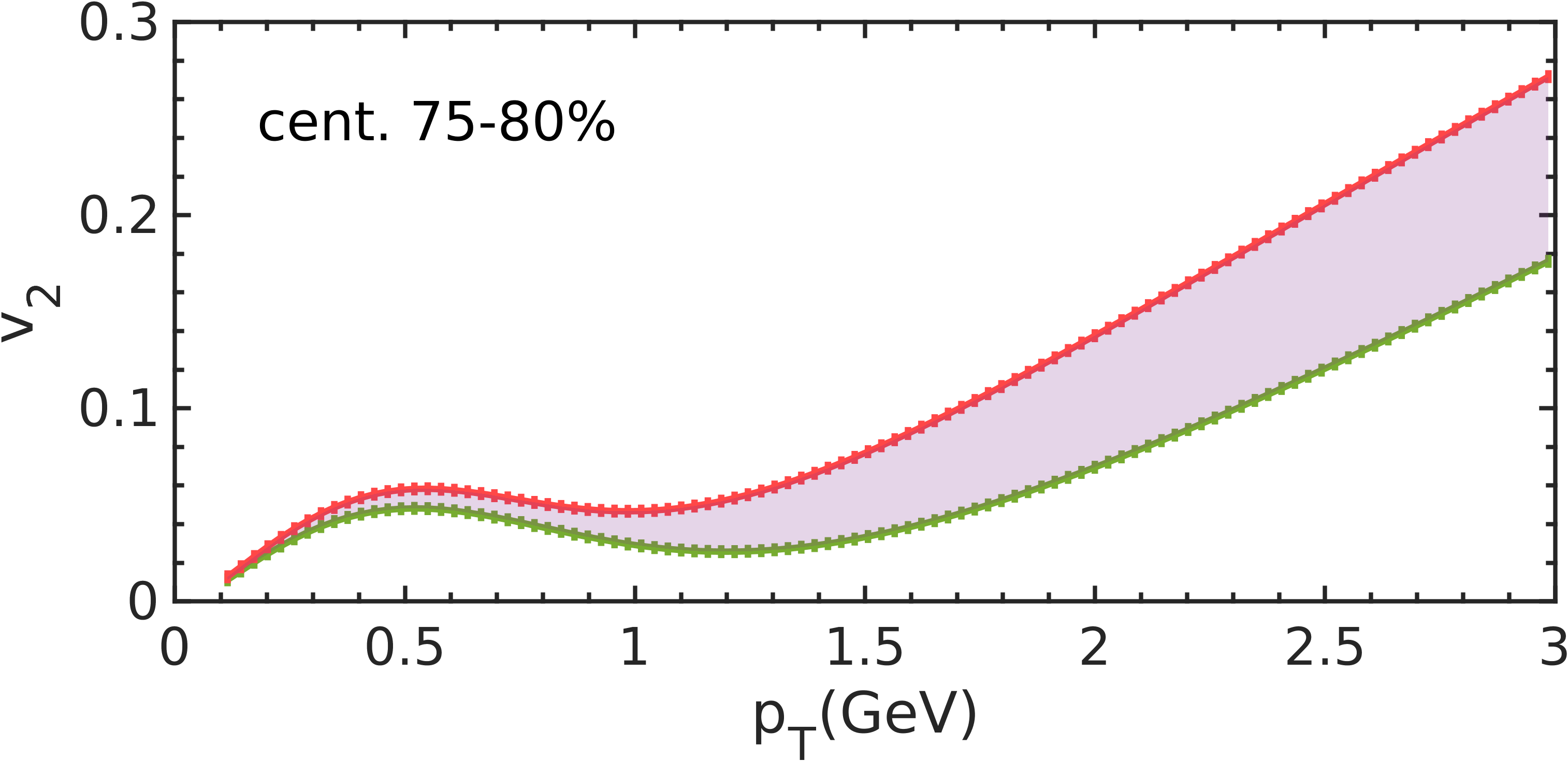}
\includegraphics[scale=0.22]{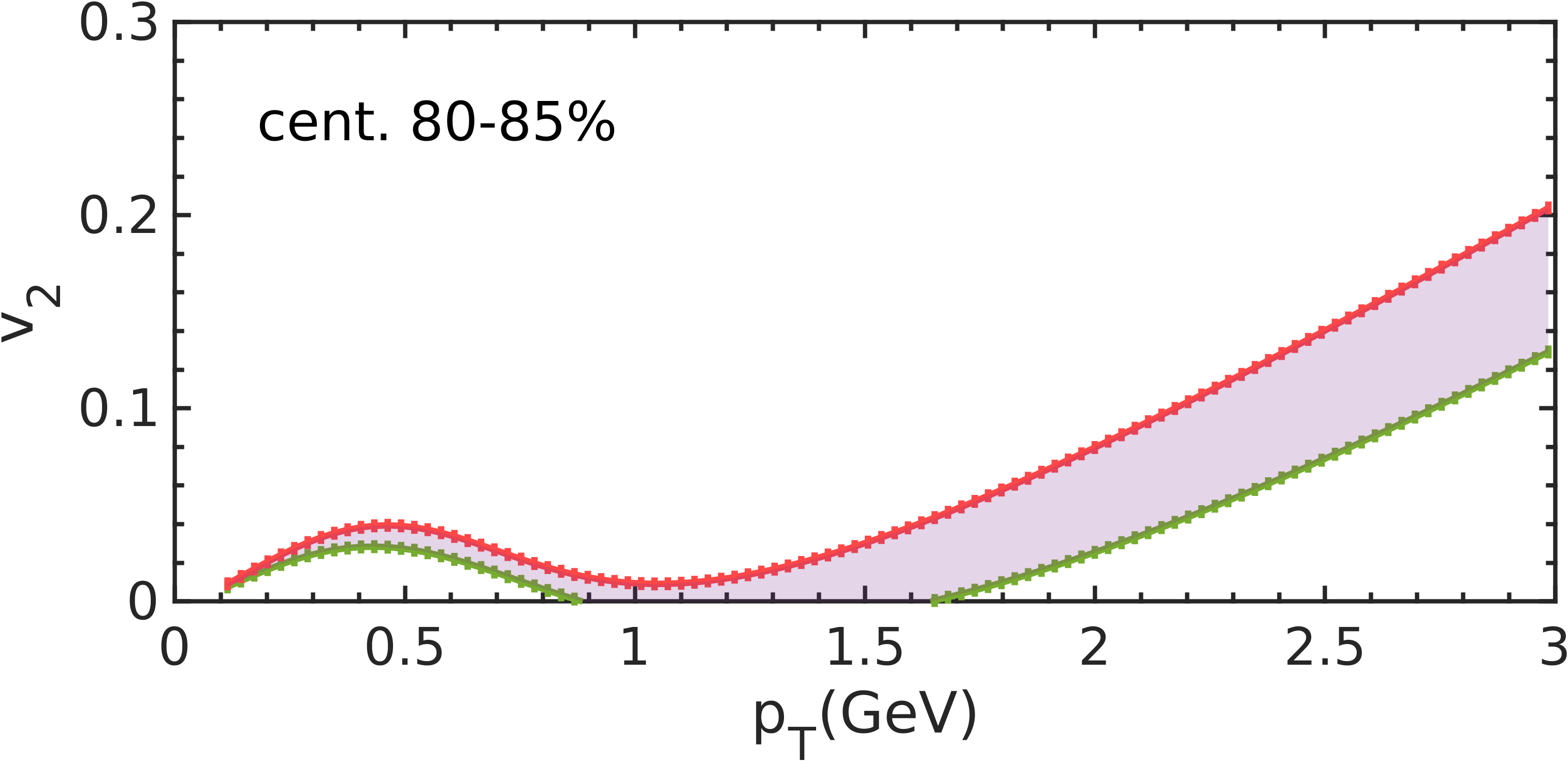}
\includegraphics[scale=0.22]{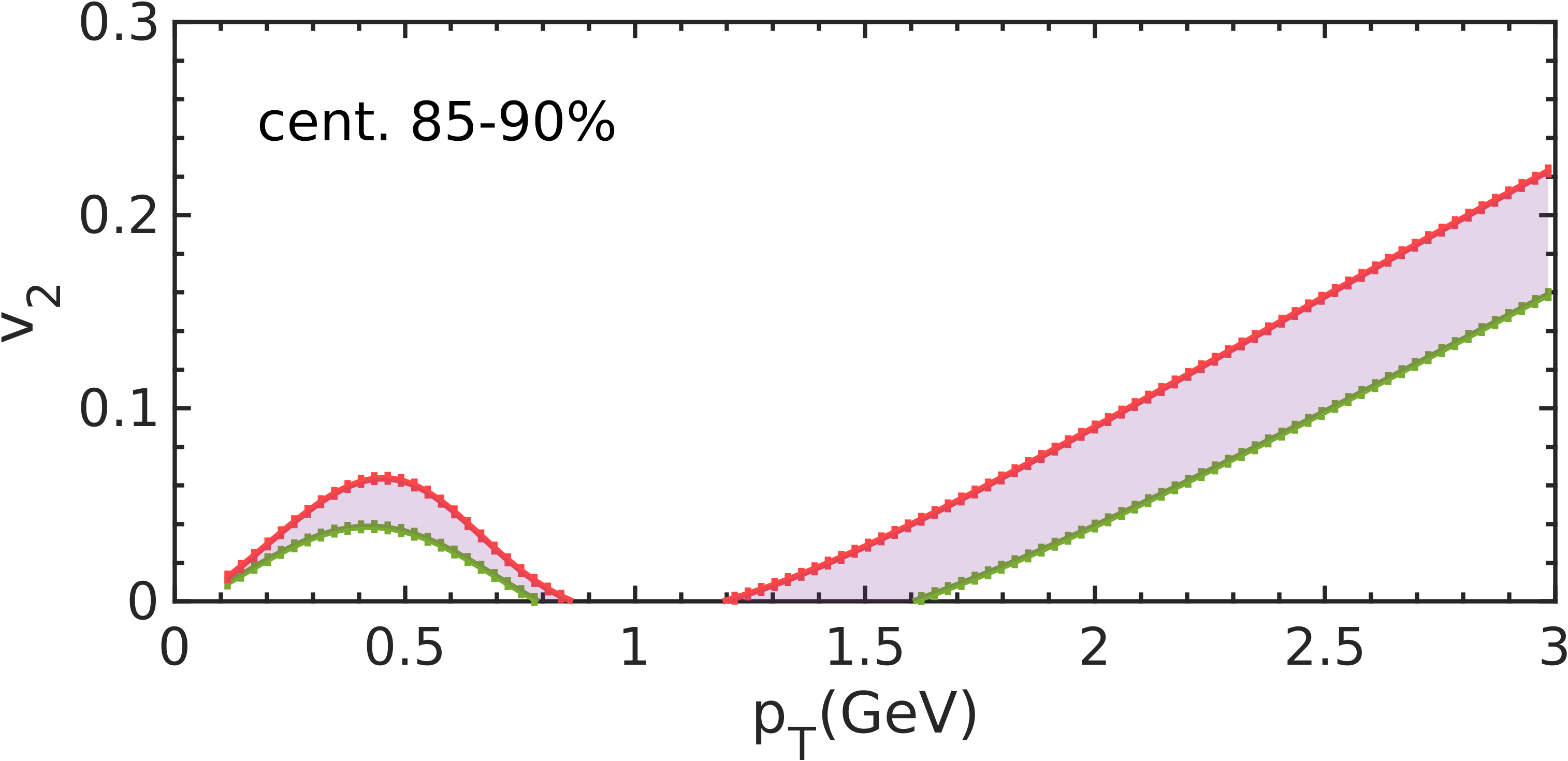}
\includegraphics[scale=0.22]{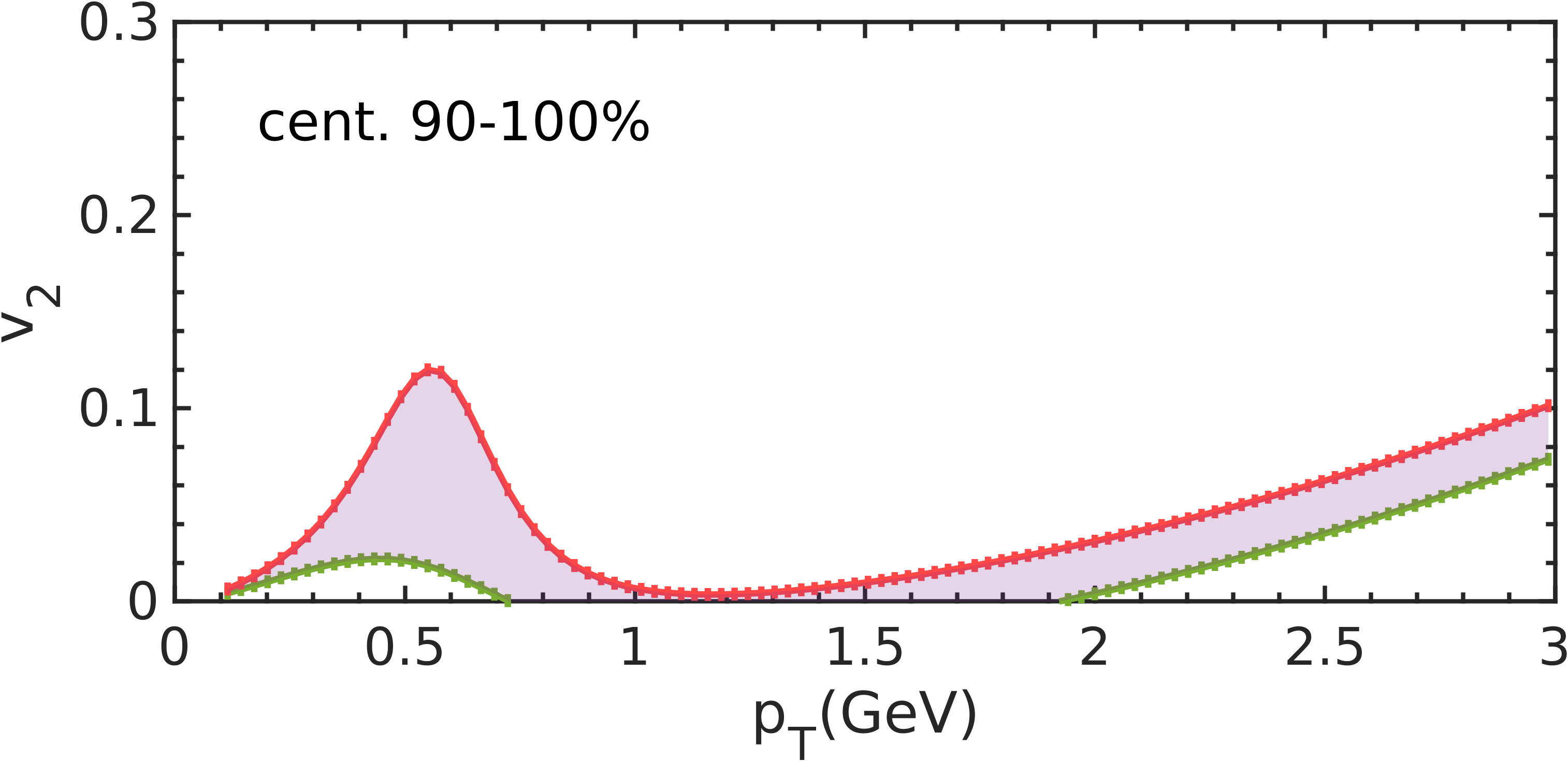}
\caption{Pion($\pi^+$) elliptic flow coefficient as a function of transverse momentum($p_T$) for $14$ centrality classes for Pb-Pb $2.76$ TeV collision system obtained with (IPGlasma+2Dhydro) setup along withe elliptic flow measured at ALICE\cite{EXPT_PbPb_v2_pT} (blue) for relaxation time $\tau_\pi=3\eta/sT$(green) and $8\eta/sT$(red). The shaded area (violet) highlights the difference in flow due to variation in relaxation time. See text for explanation.}
\label{fig:pT_v2_PbPb}
\end{figure*}

 \begin{figure}
\includegraphics[scale=0.7]{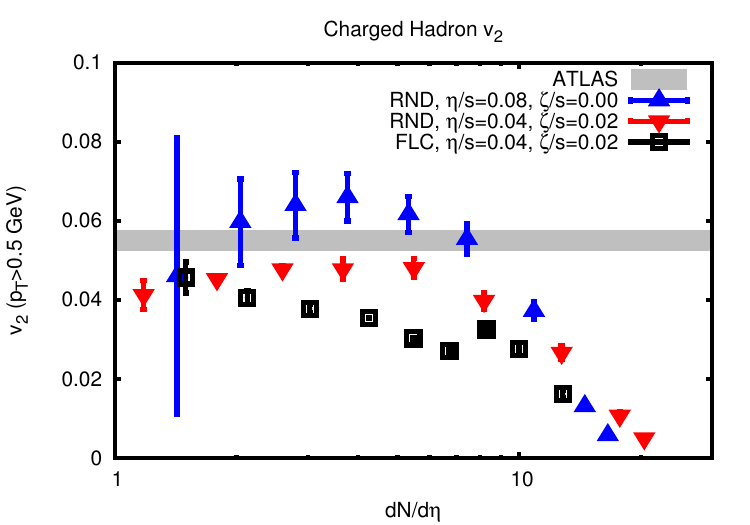}
\caption{$p_T$ integrated elliptic flow in proton-proton collision at $7$ TeV produced using SONIC model, as a function of multiplicity pseudorapidity spectra for the mentioned values of $\eta/s$ and $\zeta/s$. For $\eta/s=0.08$ and $\zeta/s=0$ (blue), elliptic flow has errorbar due to variation in shear relaxation or non-hydro mode decay time, which increases in size for decreasing $dN/d\eta$. Plot taken from~\cite{2017roma_app_hydro}.}
\label{fig:roma_fig2}
\end{figure}

\begin{figure*}
\includegraphics[scale=0.26]{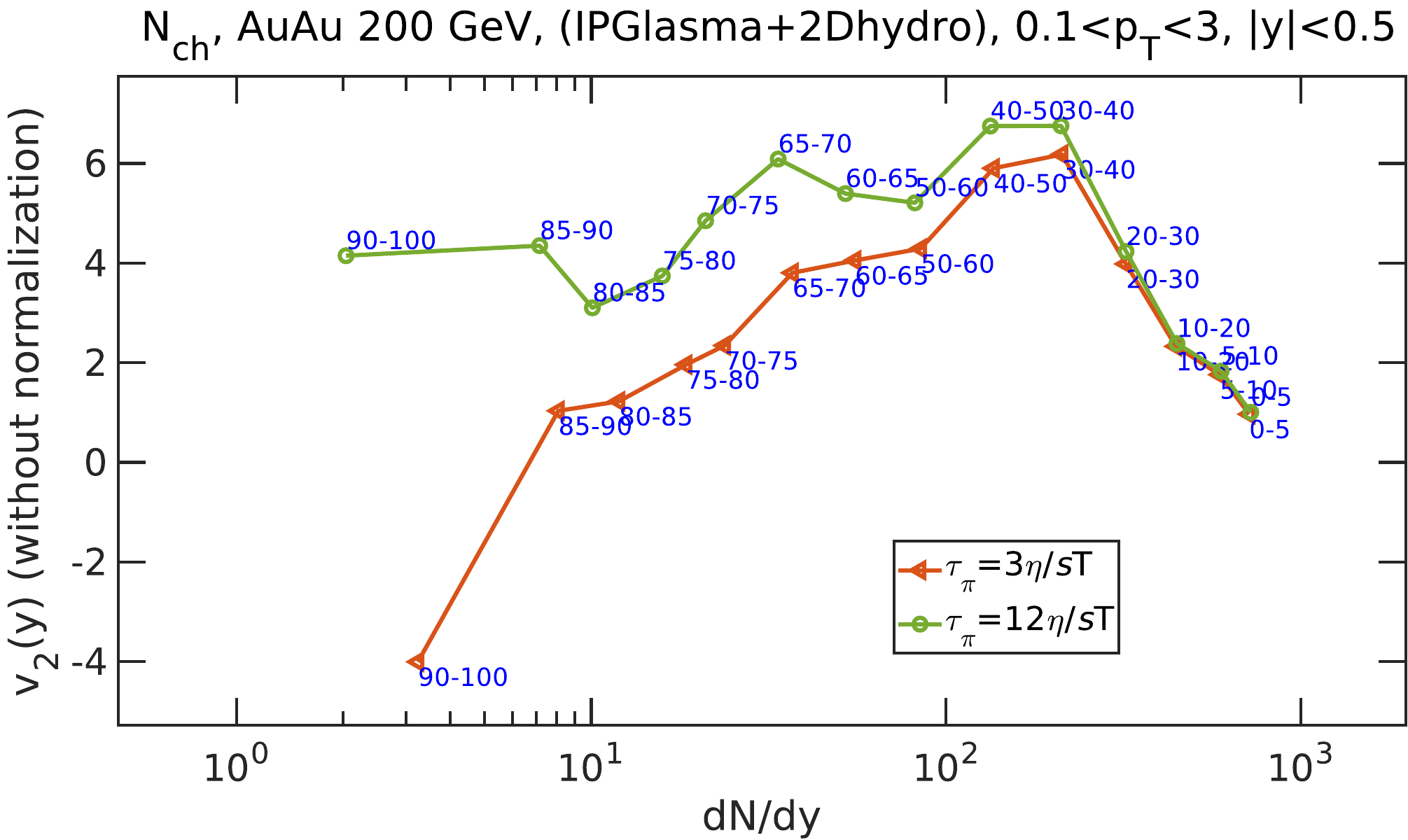}
\includegraphics[scale=0.26]{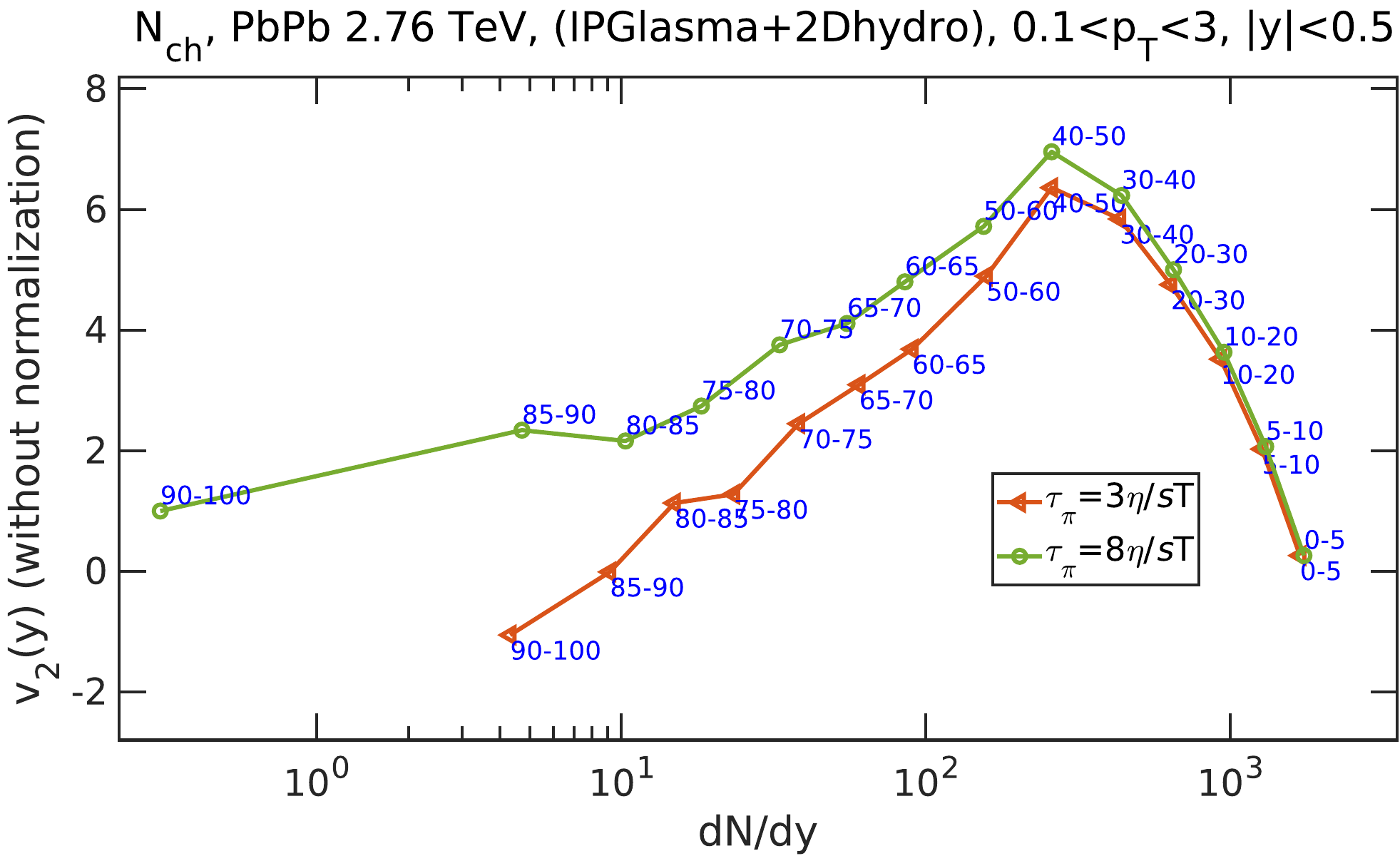}
\caption{Un-normalized $p_T$ integrated elliptic flow of charged particles as a function of $N_{ch}$ rapidity density for AuAu $200$ GeV (left) and Pb-Pb $2.76$ TeV (right) plotted for the two mentioned relaxation times. Data points are labelled by the centrality values. The separation between the two curves is better seen for un-normalized elliptic flow than for normalized one shown below.}
\label{fig:v2_vs_rap_spec2}
\end{figure*}

\begin{figure*}
\includegraphics[scale=0.26]{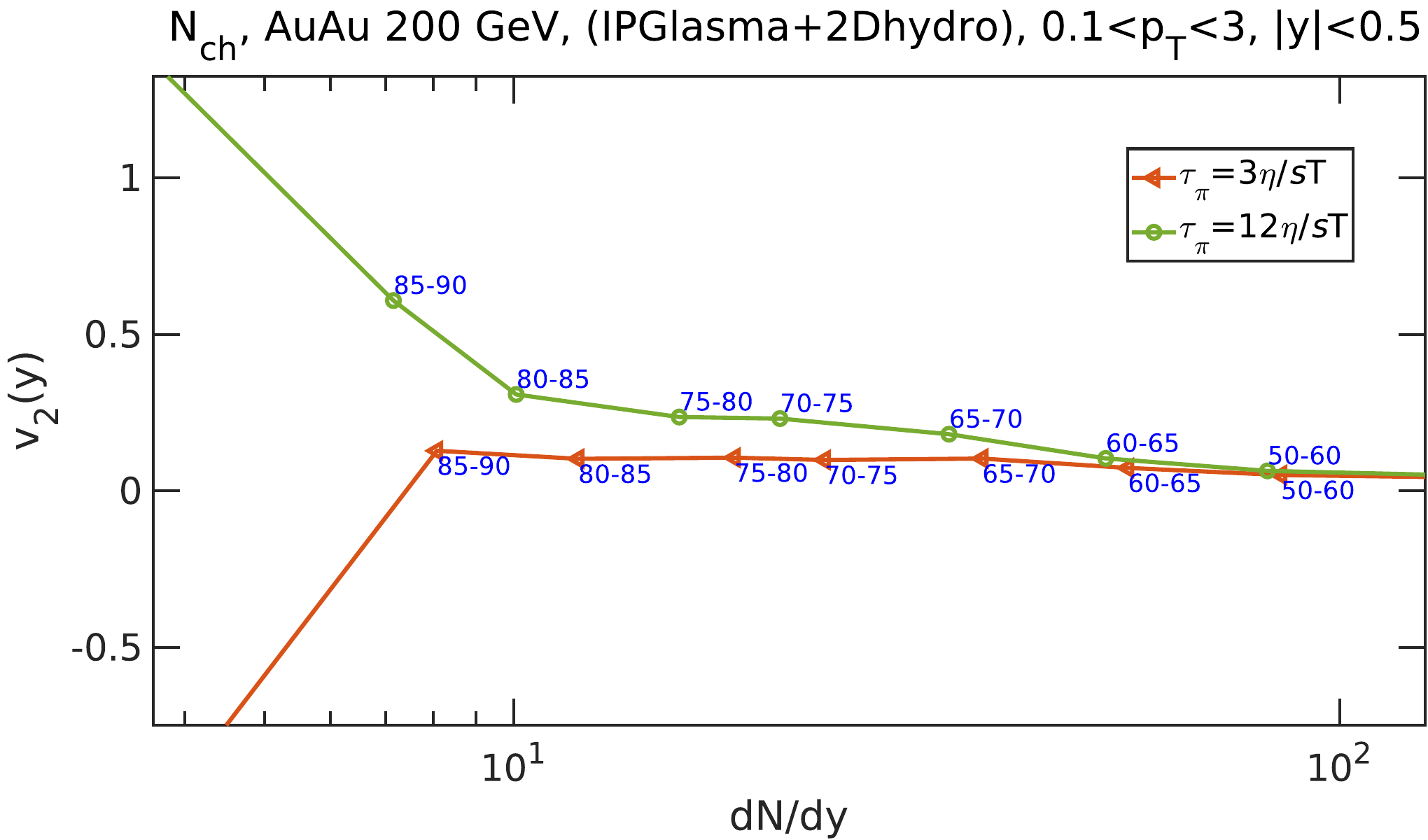}
\includegraphics[scale=0.26]{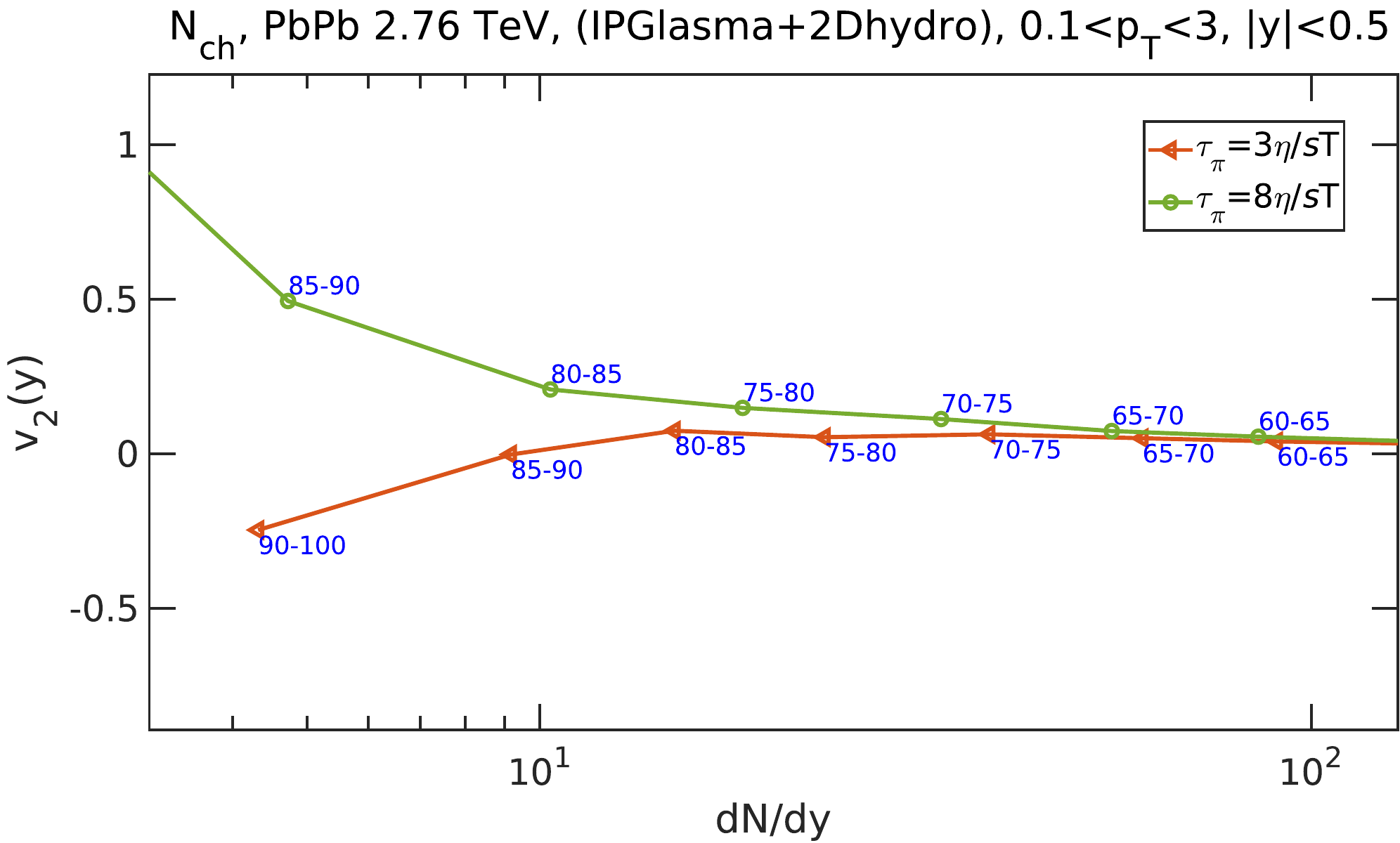}
\caption{$p_T$ integrated elliptic flow of charged particles as a function of $N_{ch}$ rapidity density for Au-Au $200$ GeV (left) and Pb-Pb $2.76$ TeV (right) plotted for the two mentioned relaxation times. Data points are labelled by the centrality values. See text for explanation.}
\label{fig:v2_vs_rap_spec1}
\end{figure*}

The second order viscous hydrodynamics used for this study is a publicly available code\footnote{http://theory.fi.infn.it/echoqgp/index.php}, ECHO-QGP~\cite{ECHOQGP2013,rolando2014ECHO}, based on MIS theory. It could be used in either $(2+1)$-D or $(3+1)$-D settings and has been utilized for bulk medium evolution in quarkonia suppression study\cite{bot_study}. Spacetime evolution of all $T^{\mu\nu}$ components could be extracted at the output. 

A tabular lattice QCD equation of state by Wuppertal-Budapest collaboration~\cite{WB2010} has been utilized. In this equation of state, the values for energy density($\epsilon$), speed of sound($c_s$) and pressure($P$) are available starting with the temperature of $100$ MeV. In order to get values below this temperature we spline interpolated temperature dependencies of quantities mentioned above with the corresponding values from hadron resonance gas model~\cite{HRG2010}. Dissipative corrections to the energy momentum tensor in ECHO-QGP are introduced in the same way as stated in Eq.(\ref{dissi_tensor}). Here the evolution of shear part of the viscous stress tensor is given by~\cite{ECHOQGP2013};
\bqa
\pi^{\mu\nu} =&& -\eta \Bigg( 2\sigma^{\mu\nu} + \frac{4}{3} \frac{\tau_\pi}{\eta} d_\mu u^\mu \pi^{\mu\nu} + \frac{\tau_\pi}{\eta} \Delta^\mu_\alpha \Delta^\nu_\beta D \pi^{\alpha\beta}  \nonumber\\
&&+  \frac{\lambda_0}{\eta} \tau_\pi ( \pi^{\mu\lambda} \Omega^\nu_\lambda + \pi^{\nu\lambda} \Omega^\mu_\lambda ) \Bigg).
\eqa
Here, $\lambda_0$ is a scalar coefficient and $\Omega$ is a traceless, anti-symmetric, transverse vorticity tensor.
$d_\mu$ is the covariant derivative given by $d_\mu u^\nu  = \partial_\mu u^\nu + \Gamma^\nu_{\beta\mu} u^\beta$, where $\Gamma^\nu_{\beta\mu}$ are the Christoffel symbols. $D = u^\mu d_\mu$, is the comoving time derivative.
The evolution of the bulk part of viscous stress tensor is given by;
\bqa
\Pi = -\zeta \Bigg( d_\mu u^\mu + \frac{\tau_\Pi}{\zeta} u^\alpha d_\alpha \Pi + \frac{4}{3} \frac{\tau_\Pi}{\zeta} \Pi \; d_\mu u^\mu \Bigg).
\eqa
The values of the transport coefficients, $\tau_\Pi$, $\lambda_0$, $\tau_\pi$, $\eta$, $\zeta$ are required for solving the above two equations, which is obtained from microscopic theory approach to hydrodynamics.
$\tau_\Pi$ is the bulk viscosity relaxation time, which represents how quickly the above $2^{nd}$ order form of bulk pressure relaxes to its leading-order form in Eq.(\ref{first_order}). The above two equations are derived under the metric signature choice of $(-1,+1,+1,+1)$.

\subsection{{\label{parameters}}Input parameters}
The form of relaxation time has been worked out for hydrodynamics beginning from numerous microscopic theories e.g., Boltzmann theory in the relativistic limit~\cite{israel1979,Roma2006_relax1}, weakly coupled QCD~\cite{2009WeaklyQCD_relax5} and AdS/CFT~\cite{2007ADS_CFT_relax3,BRSSS2008,2008conformal_relax4}. In ECHO-QGP, the relaxation time is introduced as;

\begin{equation}
\label{coe}
\tau_\pi = \tau_{coe} \frac{\eta}{sT}.
\end{equation}

The coefficient, $\tau_{coe}$ here controls the magnitude of shear relaxation time in viscous hydrodynamics. In Sec. (\ref{sec:results}), we see the consequence of varying this parameter on elliptic flow coefficients for Pb-Pb and Au-Au collisions. A transverse distribution of participating nucleons could serve as an initial condition for hydrodynamics. ECHO-QGP has an optical Glauber model as its default initial condition which assumes independent linear trajectories of nucleons in nuclei that are distributed according to Wood-Saxon distribution~\cite{2001Glauber,2007glauber}. Wood-Saxon distribution has a smooth plateau for the nucleus which decays softly towards the edges. Even though the Glauber model does not involve early stage dynamics and fluctuations of any kind, it is still a good approximation nonetheless.

IPGlasma~\cite{schenke2012IPGlasma,IPGlasma} is a more realistic initial condition that includes the dynamics beginning from the moment of collision. It is based on the color glass condensate framework. The wavefunction of a nucleus or hadron at high energy could be explained with the effective theory of color glass condensate~\cite{1994_CGC,gelis2010color}.
In IPGlasma model, the color charges inside the nucleons are Gaussian sampled and are taken as the source for gluon fields, which are then evolved using classical Yang-Mills equations~\cite{schenke2012IPGlasma}. We have used the publicly available\footnote{https://github.com/schenke/ipglasma} IPGlasma model that describes a boost invariant (2+1)-D initial state. The energy density in the transverse plane at $\tau_{start}$ = $0.2$ fm/c for Pb-Pb collision and $\tau_{start}$ = $0.6$ fm/c for Au-Au collision has been taken as an input for ECHO-QGP. Fig.(\ref{fig:ed}) shows the initial energy densities for $14$ centralities as a function of transverse coordinates for Au-Au collision. We ran the [IPGlasma initial condition + ECHO-QGP hydrodynamics] framework for $14$ centrality values, with more values near peripheral collisions.

The distribution of the nucleons in the nucleus and the distribution of color charge inside nucleons are the key sources of initial state fluctuations in each collision event. Observables in collider experiments are averaged over a large number of collision events, to account for this event-by-event fluctuation. For both Au-Au and Pb-Pb collision systems, we produce an initial state with $400$ different sets of nucleon positions that are then combined into one.
The total inelastic nucleon-nucleon cross section is set to $61.8$ mb for Pb-Pb system and $42$ mb for Au-Au system in both, IPGlasma and in hydrodynamics, taken from Monte Carlo Glauber analysis~\cite{PbPb_cent}. Shear viscosity to entropy density ratio($\eta/s$) is taken as a constant, $0.1$ ($\approx$ $1.25 \times \frac{1}{4\pi}$)~\cite{big_picture}, which is above the theoretical minimum KSS limit~\cite{KSS_limit}. Bulk viscosity has not been included in this study. The pseudo-critical temperature, at which quarks to hadron phase transition occur, has been calculated by various lattice QCD collaborations, is an input parameter. It is set to the recently calculated value of $156$ MeV~\cite{Tpc_lQCD}. Chemical freezeout is a point at which the inelastic scatterings cease to exist between produced hadrons. This point is decided by the temperature, which in the present model is fixed at $150$ MeV~\cite{chemical_freeze}.

\subsection{\label{fix_parameters} Fixing centrality parameters}

Fig. \ref{fig:pT_spectra} shows the $p_T$ spectra of pions($\pi^+$) produced for the two mentioned collision systems along with corresponding experimentally measured $p_T$ spectra. The generated spectra adequately comply with experimental values only in low $p_T$ regime, where the hydrodynamic mode operates.
The energy density profile plotted as a function of transverse coordinate from IPGlasma had to be scaled before being used in hydrodynamics. Fig. \ref{fig:ed} shows this scaled energy density distribution. This fixed the energy density scaling parameter such that the produced $p_T$ spectra and the maxima of rapidity spectra($dN/dy$) at each centrality matches with the corresponding experimental measured data for both collision systems.

Fig. \ref{fig:dNdy_Npart} shows rapidity spectra normalized to $N_{part}/2$ as a function of $N_{part}$. In addition to energy density scaling, the rapidity spectra had to be scaled to match with experimental results as shown in Fig. \ref{fig:dNdy_Npart}. For the Au-Au system, charged particle normalized rapidity spectra were scaled up by a factor of $2$, whereas for Pb-Pb system, this scaling was $6$, and the corresponding scaling used for pions was $3.6$.
We chose a centrality range spaced by 5\% in peripheral collisions except for the last centrality class, 90-100\%. The impact parameter and $N_{part}$ values for all of these centrality ranges are taken from a Monte Carlo Glauber analysis~\cite{PbPb_cent}. The reason for taking more values towards the peripheral side was to capture fine variations of flow for decreasing $dN/dy$ as could be seen in Fig. \ref{fig:v2_vs_rap_spec2} in Sec. \ref{sec:results}.
However there was no experimental reference to set parameters for these in-between centrality values for $p_T$-spectra and $dN/dy$ vs $N_{part}$ plot. Hence, we selected two values around each experimental centrality point starting from 60\% as could be seen in $dN/dy$ vs $N_{part}$ plot (Fig. \ref{fig:dNdy_Npart}). There was no experimental point at 90-100\% so we settled with just one extrapolated value which follows the trend of data.
The blue labels on data points in Fig. \ref{fig:dNdy_Npart} are the mid centrality value of that data point.
For calculating observables for charged particles, we have added the corresponding values for the pions($\pi^+$+$\pi^-$), kaons($K^+$+$K^-$) and protons($p^+$+$p^-$) since these are abundantly produced species in high energy collisions.
Momentum space eccentricity which is the precursor of elliptic flow can be calculated in terms of $T^{\mu\nu}$ components as:

\begin{equation}
e_p\equiv  \frac{\int d^2x_\perp ( T^{xx}-T^{yy} )}{\int d^2x_\perp ( T^{xx}+T^{yy} )}
\end{equation}

ECHO-QGP calculates this quantity for ideal hydrodynamic case, which takes the form:

\begin{equation}
e_p \equiv \frac{\int d^2x_\perp (\epsilon+P) \left(u^x u^x-u^y u^y\right)}{\int d^2x_\perp [ (\epsilon+P) \left(u^x u^x+u^y u^y\right)+2P ]}
\end{equation}
To generate momentum eccentricity for the viscous case, we have modified the above expression by adding viscous component term, ($\pi^{xx} + \pi^{yy}$) to the integrand in both numerator and denominator.

Fig. \ref{fig:ecc} shows spatial eccentricity($\epsilon_c$) and momentum space eccentricity($\epsilon_p$) for Au-Au and Pb-Pb collision, generated at $50-60$\% centrality for the two mentioned shear relaxation times. Momentum anisotropy quantified by momentum eccentricity increases at the expense of spatial anisotropy quantified by spatial eccentricities along the evolution\cite{2008hydro}. The variation in non-hydrodynamic mode decay time seems to have negligible effect on spatial eccentricity. The distinguishing feature between the two systems is that the early time $\epsilon_c$ for Pb-Pb decreases more rapidly than that for Au-Au collisions.
Below the pseudo-critical temperature, hadronic picture should emerge. Particles of various species are assigned momentum according to Cooper–Frye scheme~\cite{Cooper_frye}. The resulting momentum spectrum is then used to calculate the elliptic flow, $v_2 = \langle \cos[2({\phi - \Psi_{\text{RP}}})]\rangle$, where $\Psi_{\text{RP}}$ is the reaction plane angle which acts as a reference plane and $\phi$ is the transverse plane angle for a given particle with respect to the reaction plane.

Fig. \ref{fig:avg_pT} shows the average transverse momentum evolution as a function of centrality.
Results for the two values of shear relaxation time have been plotted and compared with experimental values for pions. We notice, that the model show agreement with experimental values for most of the centrality classes apart from the peripheral ones.
The values for Pb-Pb collisions had to be scaled up by a factor of $1.3$. This could be due to underproduction of hadrons in the hydrodynamics, because the multiplicity has been used as the weight factor for calculating mean $p_T$.

\section{\label{sec:results}Flow Results and Discussion}

Romatschke \cite{2017roma_app_hydro} has put forth a quantitative test for applicability of hydrodynamics by checking the sensitivity of certain observables(like elliptic flow) to the non-hydrodynamic mode. The idea is that hydrodynamics can be used to describe a system if the non-hydrodynamic mode is sub-dominant and there exists a local rest frame. With QCD as the microscopic theory, approximate transverse momentum range of hydrodynamic mode is 3 to 7 GeV. Fig. \ref{fig:roma_fig1} illustrates this $p_T$-range where hydro and non-hydro modes operate.

This is what we have tried checking for Au-Au 200 GeV in Fig. \ref{fig:pT_v2_AuAu} and for Pb-Pb 2.76 TeV in Fig. \ref{fig:pT_v2_PbPb}. Peripheral collisions are the system of interest, but experimentally measured anisotropic flow results are only available upto 50-60\% centrality class. We hence presented the results for the complete centrality range.
In Fig. \ref{fig:pT_v2_AuAu}, for 0-5\%, 5-10\% and 10-20\% centralities, we see no separation between elliptic flow curves for non-hydrodynamic mode decay times, $\tau_\pi=3\eta/sT$ and $12\eta/sT$. From 20-30\% centrality class onwards we notice the separation between these two flow curves to be increasing. Experimental data has been plotted just for reference that show our results are quite close to experimentally measured flow results. The important point to notice is that along increasing centrality, the point at which the two flow curves separate shift towards lower $p_T$ values. Which means that with increasing centrality and decreasing system size, the hydrodynamic mode is shrinking and non-hydrodynamic mode is getting dominant. Hence in a way we are witnessing limit of applicability of low-order hydrodynamics for decreasing system size at constant collisional energy(here, 200 GeV).

Fig. \ref{fig:pT_v2_PbPb} shows $p_T$ dependence of pion($\pi^+$) elliptic flow  with complete centrality range for $\tau_\pi=3\eta/sT$ and $8\eta/sT$. We notice all the structures mentioned above for Au-Au, 200 GeV system. We notice a better match between produced elliptic flow and experimental data 10-20\% onwards.
We chose pions for this analysis because they are the lightest of particle species produced and hence adequately represents the bulk medium.
One additional point to notice is, for $10-20$\% centrality in Au-Au collisions and classes $0-5$\%, $5-10$\% in Pb-Pb collision system, our model fails to reproduce the measured elliptic flow data.

Fig. \ref{fig:roma_fig2} depicts the criteria suggested by Romatschke to check the applicability of hydrodynamics. This  figure shows charged particles elliptic flow as a function of multiplicity pseudorapidity density for proton-proton collision. The errorbar depicts the abrupt change in flow due to variation in non-hydrodynamic mode decay time. This abrupt change in elliptic flow is indicative of breakdown of hydrodynamics, and it is seemingly happening at roughly $dN/d\eta < 2$ in Fig. \ref{fig:roma_fig2}. We tried checking this feature in our (IPGlasma+2Dhydro) analysis as shown in Figs. \ref{fig:v2_vs_rap_spec1} and \ref{fig:v2_vs_rap_spec2}.

Fig. \ref{fig:v2_vs_rap_spec2} presents the un-normalized $p_T$ integrated elliptic flow as a function of multiplicity rapidity density($dN/dy$). The data points from our analysis are labelled by the centrality class in order to track the point at which flow changes abruptly between the relaxation time curves. This is why we selected more centrality points in peripheral collision side.
We notice a steady increase in separation between the two flow curves for both Au-Au and Pb-Pb system which is in reasonably close agreement with Romatschke's work.

We also notice that the two relaxation time flow curves of same centrality do not have same multiplicity rapidity density value (the x co-ordinate). This would mean, that for an increase in relaxation time, flow shifts to a lower multiplicity value. We also notice that the flow for $\tau_\pi=3\eta/sT$ for both, Au-Au and Pb-Pb systems, acquire negative values, which is also apparent from the elliptic flow for 90-100\% centrality class in Figs. \ref{fig:pT_v2_PbPb} and \ref{fig:pT_v2_AuAu}.

Fig. \ref{fig:v2_vs_rap_spec1} shows normalized $p_T$ integrated elliptic flow as a function of charged particle multiplicity rapidity density($dN/dy$) for peripheral collisions. We clearly notice the sudden increase in separation of flow curves for the two mentioned relaxation times for both the collision systems. But we don't have a centrality resolution good enough to decide the onset of hydrodynamization. A approximate limit we can deduce from Fig. \ref{fig:v2_vs_rap_spec1} is $dN/dy\approx10$ which is quite larger than the prediction of $dN/d\eta < 2$ ~\cite{roma2017_hydro,heinz2019_how_the_heck}. However if the hadron resonance gas to de-confined quarks transition in high temperature regime is a crossover, we expect to find a region where analysis would be indecisive like what Aleksi Kurkela \textit{et al}. obtained ~\cite{kurkela2019flow,kurkela2019opacity}. The problem lies in the absence of experimental reference data to set the scaling parameter of IPGlasma for such high centrality classes.

\section{\label{sec:conclu} Conclusion and outlook}

In this study we analyze the non-hydrodynamic mode in an attempt to find the onset of hydrodynamization in peripheral collision system of Au-Au and Pb-Pb at $200$ GeV and $2.76$ TeV center of mass per energy nucleon, respectively. We use the energy density profile from \textit{color glass condensate} based IPGlasma model as the initial condition in 2D ECHO-QGP which is a $2^{nd}$ order viscous hydrodynamic code based on MIS theory. $p_T$ spectra and multiplicity rapidity density($dN/dy/(N_{part}/2)$) as a function of $N_{part}$ is used to constrain the centrality scaling parameter of IPGlasma. Mean $p_T$ as a function of centrality, evolution of spatial and momentum eccentricity has also been generated for both the systems. The shear viscosity to entropy density ratio is set as $\eta/s=0.1$ and bulk viscosity has not been considered in this work. We study the variation in the strength of non-hydrodynamic mode through the shear relaxation time, whose value is set to $(3-12)\eta/sT$ for Au-Au system and $(3-8)\eta/sT$ for Pb-Pb system. Elliptic flow generated as a function of $p_T$ is compared with $2^{nd}$ anisotropic flow coefficient from experiments for the above respective values of relaxation time, for all of the $14$ centrality classes. Normalized and un-normalized $p_T$ integrated elliptic flow has been studied as a function of multiplicity rapidity density in peripheral collisions especially. We found the following:
\begin{itemize}
\item From $p_T$ dependence of elliptic flow across centralities for Au-Au in Fig. \ref{fig:pT_v2_AuAu} and for Pb-Pb in Fig. \ref{fig:pT_v2_PbPb}, we found that the shear relaxation time does control the non-hydrodynamic mode of the system as predicted by P. Romatschke. This inference was guided by the observation that the point after which the flow for the two relaxation times separate sharply from each other, shifts to lower $p_T$ values for increasing centrality classes (or decreasing system size at a constant energy of collision). 
\item We later attempted testing the onset of hydrodynamization from charged particle multiplicity rapidity density dependence of $p_T$ integrated elliptic flow. We did notice an abrupt increase in flow for decreasing system size or number of participants, indicating increased dominance of non-hydrodynamic mode and simultaneous breakdown of hydrodynamic description. However we could not resolve the $dN/dy$ below the value of  $10$ enough to quantitatively decide the onset point.
\item We found a good agreement between the generated $p_T$ dependence of elliptic flow results and the measured flow data from PHENIX and ALICE Collaborations for Au-Au and Pb-Pb systems, respectively, except near most centrality of 10-20\% class for Au-Au collisions and and of 0-5\% and 5-10\% class for Pb-Pb collisions.
\end{itemize}

There is significant scope for improving this framework further by including an after-burner stage that will incorporate hadron resonance decays and scattering which could affect the generated flow\cite{bass1998microscopic}. It will be interesting to compare lowest fluid size from other methods in the future work.
 The initial state involvement could also be improved by using more components of $T^{\mu\nu}$ in hydrodynamics \cite{AMPT+hydro,schenke2020hybrid}. One can also switch to $3-$D IPGlasma initial condition~\cite{3D_IPGlasma}. Bulk viscosity has been kept zero in this study. But it does play significant role in evolution~\cite{2015bulk}. The relaxation times for bulk viscosity could be independently analyzed. It will be interesting to see if elliptic flows of different particle species diverge for decreasing rapidity spectra at different points. If they do so, it would support the idea of multiple-fluid scenario in heavy ion collision. This study could be extended to small and lower energy system where the net-baryon potential is non-zero, for which particle current conservation should be included\cite{du2020_BEShydro}. In a recent study, Plumberg \textit{et al}. \cite{causality_violation}, have conducted a causality analysis of each fluid cell of hydrodynamics for its complete evolution. They found causality being violated of non-hyperbolic($v^2<0$) and superluminal($v^2 > c^2$) type at early times in evolution. This violation is significantly reduced if a pre-equilibrium stage like K$\o$MP$\o$ST \cite{kompost} is used. It will be interesting to see the repercussions of such a study on onset of hydrodynamization.

\begin{acknowledgements}
We are thankful to Gabriele Inghirami for clearing our doubts and helping at numerous times in using the viscous hydrodynamics code. We are also grateful to Chun Shen, Paul Romatschke and Rajeev Bhalerao for clearing our doubts about flow. Nikhil Hatwar would like to thank BITS-Pilani for the financial support.
\end{acknowledgements}

\bibliography{v12_6Oct22}
\end{document}